%% file: HA-general.tex
\documentclass{elsarticle}

\usepackage{enumitem} 
\setlist{noitemsep,topsep=0pt,parsep=0pt,partopsep=0pt,listparindent=\parindent} 

\usepackage[margin=1cm]{caption} 
\usepackage{subfig}
\usepackage{tikz}
\usetikzlibrary{arrows,backgrounds,fit}

\tikzset{every fit/.append style=text badly centered}

\usepackage{tyson}

\usepackage[bookmarks=true,hypertexnames=false]{hyperref}
\hypersetup{colorlinks=true, citecolor=blue, linkcolor=red, urlcolor=blue}

\newcommand{\Holant}{\operatorname{Holant}}
\newcommand{\holant}[2]{\Holant\left(#1\mid #2\right)}

\newcommand{\arity}{\operatorname{arity}}

\newcommand{\StabA}{\operatorname{Stab}(\mathscr{A})}

\newcommand{\StabP}{\operatorname{Stab}(\mathscr{P})}

\newenvironment{remark}{\medskip{\bfseries \noindent Remark:}}{\par\medskip}{\par\medskip}

\def\scale{0.6}
\def\nodeDist{1.4cm}
\tikzstyle{internal} = [draw, fill, shape=circle]
\tikzstyle{external} = [shape=circle]
\tikzstyle{square}   = [draw, fill, rectangle]

\begin{document}

\title{Holographic Algorithms Beyond Matchgates\tnoteref{conf}}
\tnotetext[conf]{A preliminary version has appeared in ICALP 2014 \cite{CGW14}.}

\cortext[cor1]{Corresponding author}

\author[uwm]{Jin-Yi Cai}
\ead{jyc@cs.wisc.edu}
\address[uwm]{University of Wisconsin - Madison, 1210 West Dayton Street, Madison, WI 53706, US}

\author[edi]{Heng Guo\corref{cor1}}
\ead{hguo@inf.ed.ac.uk}
\address[edi]{University of Edinburgh, Informatics Forum, 10 Crichton Street, Edinburgh, EH8 9AB, UK}

\author[blo]{Tyson Williams}
\ead{tdw@cs.wisc.edu}
\address[blo]{Blocher Consulting, 206 North Randolph, Champaign, IL 61820, US}

\begin{abstract}
  Holographic algorithms introduced by Valiant are composed of two ingredients: 
  matchgates, which are gadgets realizing local constraint functions by weighted planar perfect matchings,
  and holographic reductions, which show equivalences among problems with different descriptions via certain basis transformations.
  In this paper, we replace matchgates in the paradigm above by the \emph{affine} type and the \emph{product} type constraint functions,
  which are known to be tractable in general (not necessarily planar) graphs.
  More specifically, we present polynomial-time algorithms to decide 
  if a given counting problem has a holographic reduction to another problem defined by the affine or product-type functions.
  Our algorithms also find a holographic transformation when one exists.
  We further present polynomial-time algorithms of the same decision and search problems for symmetric functions,
  where the complexity is measured in terms of the (exponentially more) succinct representations.
  The algorithm for the symmetric case also shows that the recent dichotomy theorem for Holant problems with symmetric constraints is efficiently decidable.
  Our proof techniques are mainly algebraic, e.g., using stabilizers and orbits of group actions.
\end{abstract}


\maketitle

\input{1.Introduction}

\input{2.Preliminaries}

\input{3.Problems}

\input{4.General.Affine}

\input{5.General.Product}

\input{6.1.Symmetric.Affine.Single}

\input{6.2.Symmetric.Affine.Set}

\input{7.Symmetric.Product}

\paragraph{Acknowledgements}

This research is supported by NSF CCF-1714275.
HG was also supported by a Simons Award for Graduate Students in Theoretical Computer Science from the Simons Foundation when he was a graduate student.
We thank anonymous referees for their valuable comments.

\bibliographystyle{plain}
\bibliography{bib}

\end{document}

%% file: 1.Introduction.tex
\section{Introduction}

Recently a number of complexity dichotomy theorems have been obtained for counting problems.
Typically, such dichotomy theorems assert that a vast majority of problems expressible within certain frameworks are $\SHARPP$-hard,
however an intricate subset manages to escape this fate.
These exceptions exhibit some rich mathematical structure,
leading to polynomial-time algorithms.
Holographic reductions and algorithms, introduced by Valiant \cite{Val08}, 
play key roles in many recent dichotomy theorems~\cite{CK12, CLX17, CLX11a, CKW12, HL16, CLX13, CGW16, GW13}.
Indeed, many interesting tractable cases are solvable using holographic reductions.
This fascinating fact urges us to explore the full reach of holographic algorithms.


Valiant's holographic algorithms~\cite{Val08, Val06} have two main ingredients.
The first is to encode computation in planar graphs via gadget construction, called matchgates~\cite{Val02a, Val02b, CCL09, CL11a, CG14}.
The result of the computation is then obtained by counting the number of perfect matchings in a related planar graph,
which can be done in polynomial time by Kasteleyn's (a.k.a.~the FKT) algorithm~\cite{Kas61, TF61, Kas67}.
The second one is the notion of holographic transformations/reductions,
which show equivalences of problems with different descriptions via basis transformations.
Thus, in order to apply the holographic algorithm,
one must find a suitable holographic transformation along with matchgates realizing the desired constraint functions.
This procedure has been made algorithmic~\cite{CCL09, CL11a}.

In this paper, we replace matchgates in the paradigm above by the \emph{affine} type or the \emph{product} type constraint functions,
both of which are known to be tractable over general (i.e.~not necessarily planar) graphs~\cite{CLX14}.
We present polynomial-time algorithms to decide 
if a given counting problem has a holographic reduction to another problem defined by affine or product-type functions.
Our algorithm also finds a holographic reduction when one exists.
Although, conceptually, we do not add new tractable cases,
the task of finding these transformations is often non-trivial.
For example, generalized Fibonacci gates \cite{CLX11c} are the same as the product-type via transformations,
but at first glance, the former look much more complicated than the latter.

To formally state the results, we briefly introduce some notation.
The counting problems we consider are those expressible as a Holant problem~\cite{CLX11c, CLX12, CLX11d, CLX14}.
A Holant problem is defined by a set $\mathcal{F}$ of constraint functions, which we call signatures, and is denoted by $\Holant(\mathcal{F})$.
An instance of $\Holant(\mathcal{F})$ is a tuple $\Omega = (G, \mathcal{F}, \pi)$, called a signature grid,
where $G = (V,E)$ is a graph and $\pi$ labels each vertex $v \in V$ and its incident edges with some $f_v \in \mathcal{F}$ and its input variables.
Here $f_v$ maps $\{0,1\}^{\deg(v)}$ to $\mathbb{C}$, where $\deg(v)$ is the degree of $v$.
We consider all possible 0-1 edge assignments.
An assignment $\sigma$ to the edges $E$ gives an evaluation $\prod_{v \in V} f_v(\sigma |_{E(v)})$,
where $E(v)$ denotes the incident edges of $v$ and $\sigma |_{E(v)}$ denotes the restriction of $\sigma$ to $E(v)$.
The counting problem on the instance $\Omega$ is to compute
\[
 \Holant_\Omega = \sum_{\sigma : E \to \{0,1\}} \prod_{v \in V} f_v\left(\sigma |_{E(v)}\right).
\]
For example, consider the problem of counting \textsc{Perfect Matching} on $G$.
This problem corresponds to attaching the \textsc{Exact-One} function at every vertex of $G$.
The \textsc{Exact-One} function is an example of a symmetric signature,
which are functions that only depend on the Hamming weight of the input.
We denote a symmetric signature by $f = [f_0, f_1, \ldots, f_n]$ where $f_w$ is the value of $f$ on inputs of Hamming weight $w$.
For example, $[0,1,0,0]$ is the \textsc{Exact-One} function on three bits.
The output is~$1$ if and only if the input is $001$, $010$, or $100$, and the output is~$0$ otherwise.

Holant problems contain both counting constraint satisfaction problems and counting graph homomorphisms as special cases.
All three classes of problems have received considerable attention,
which has resulted in a number of dichotomy theorems
(see~\cite{Lov67, HN90, DG00, BG05, DGP07, BD07, DGJ09, BDGJR09, CC10, GGJT10, GHLX11, CCL16, CHL12, CK12, CC17, DR13, Bul13, CCL13, CLX14}).
Despite the success with \#CSP and graph homomorphisms,
the case with Holant problems is more difficult.
Recently, a dichotomy theorem for Holant problems with symmetric signatures was obtained \cite{CGW16},
but the general (i.e.~not necessarily symmetric) case has a richer and more intricate structure.
The same dichotomy for general signatures remains open.

Our first main result is an efficient procedure to decide whether a given Holant problem can be solved by affine or product-type signatures via holographic transformations.
In past classification efforts, we have been in the same situation several times, 
where one concrete problem determines the complexity of a wide range of problems.
However, the brute force way to check whether this concrete problem already belongs to known tractable classes is time-consuming.
We hope that the efficient decision procedure given here mitigates this issue, and would help the pursuit towards a general Holant dichotomy.

\begin{theorem} \label{thm:main:general}
 There is a polynomial-time algorithm to decide,
 given a finite set of signatures $\mathcal{F}$,
 whether $\Holant(\mathcal{F})$ admits a holographic algorithm based on affine or product-type signatures.
\end{theorem}

The holographic algorithms for $\Holant(\mathcal{F})$ are all polynomial time in the size of the problem input $\Omega$.
The polynomial time decision algorithm of Theorem~\ref{thm:main:general} is on another level;
it decides based on any specific set of signatures $\mathcal{F}$ whether the counting problem $\Holant(\mathcal{F})$ defined by $\mathcal{F}$ has such a holographic algorithm.

Symmetric signatures are an important special case.
Because symmetric signatures can be presented exponentially more succinctly, 
we would like the decision algorithm to be efficient when measured in terms of this succinct description.
An algorithm for this case needs to be exponentially faster than the one in Theorem~\ref{thm:main:general}.
In Theorem~\ref{thm:main:symmetric},
we present a polynomial time algorithm for the case of symmetric signatures.
The increased efficiency is based on several signature invariants under orthogonal transformations.

\begin{theorem} \label{thm:main:symmetric}
 There is a polynomial-time algorithm to decide,
 given a finite set of symmetric signatures $\mathcal{F}$ expressed in the succinct notation,
 whether $\Holant(\mathcal{F})$ admits a holographic algorithm based on affine or product-type signatures.
\end{theorem}

A dichotomy theorem classifies every set of signatures as defining either a tractable problem or an intractable problem (e.g.~$\SHARPP$-hard).
Yet it would be more useful if given a specific set of signatures,
one could decide to which case it belongs.
This is the decidability problem of a dichotomy theorem.
In~\cite{CGW16}, a dichotomy regarding symmetric complex-weighted signatures for Holant problems was proved.
However, the decidability problem was left open.
Of the five tractable cases in the dichotomy theorem, three of them are easy,
but the remaining two cases are more challenging,
which are~(1) holographic algorithms using affine signatures and~(2) holographic algorithms using product-type signatures.
As a consequence of Theorem~\ref{thm:main:symmetric}, this decidability is now proved.

\begin{corollary} \label{cor:decide:symmetric}
 The dichotomy theorem for symmetric complex-weighted Holant problems in~\cite{CGW16} is decidable in polynomial time.
\end{corollary}

Previous work on holographic algorithms focused almost exclusively on those with matchgates~\cite{Val08, Val06, CL08b, CLX17, CL11a, CL11b, GW13}.
(This has led to a misconception in the community that holographic algorithms are always based on matchgates.)
The first example of a holographic algorithm using something other than matchgates came in~\cite{CLX11c}.
These holographic algorithms use generalized Fibonacci gates.
A symmetric signature $f = [f_0, f_1, \dotsc, f_n]$ is a generalized Fibonacci gate of type $\lambda \in \mathbb{C}$
if $f_{k+2} = \lambda f_{k+1} + f_k$ holds for all $k \in \{0,1,\dotsc,n-2\}$.
The standard Fibonacci gates are of type $\lambda = 1$, in which case,
the entries of the signature satisfy the recurrence relation of the Fibonacci numbers.
The generalized Fibonacci gates were immediately put to use in a dichotomy theorem~\cite{CLX12}.
As it turned out, for nearly all values of $\lambda$,
the generalized Fibonacci gates are equivalent to product-type signatures via holographic transformations.
Our results provide a systematic way to determine such equivalences
and we hope these results help in determining the full reach of holographic algorithms.

The constraint functions we call signatures are essentially tensors.
A group of transformations acting upon these tensors yields an orbit.
Previously, in \cite{CGW16}, we have shown that it is sufficient to restrict holographic transformations to those from or related to the orthogonal group
(see Lemma~\ref{lem:affine:trans} and Lemma~\ref{lem:product:trans}).
Thus, our question can be rephrased as the following: given a tensor, 
determine whether its orbit under the orthogonal group action (or related transformations) intersects the set of affine or product-type tensors.
As showed by Theorems~\ref{thm:main:general} and~\ref{thm:main:symmetric}, 
this can be done efficiently, even for a set rather than a single tensor.
In contrast, this orbit intersection problem with the general linear group acting on two arbitrary tensors is $\NP$-hard~\cite{Kay12}.
In our setting, the actions are much more restricted and we consider an arbitrary tensor against one of the two fixed sets.
Similar orbit problems are central in geometric complexity theory~\cite{MS01}.

Our techniques are mainly algebraic.
A particularly useful insight is that an orthogonal transformation in the standard basis
is equivalent to a diagonal transformation in the $\tbmatrix{1}{1}{i}{-i}$ basis.
Since diagonal transformations are much easier to understand,
this gives us some leverage to understand orbits under orthogonal transformations.
Also, the groups of transformations that stabilize the affine and product-type signatures play important roles in our proofs.
Comparing to similar results for matchgates \cite{CL11a},
the proofs are very different in that each proof relies heavily on distinct properties of matchgates or the affine and product-type signatures.

In Section~\ref{sec:preliminaries}, we review basic notation and state previous results, many of which come from~\cite{CGW16}.
In Section~\ref{sec:problems}, we present some example problems that are tractable by holographic algorithms using affine or product-type signatures.
The proof of Theorem~\ref{thm:main:general} spans two sections.
The affine case is handled in Section~\ref{sec:general:affine} and the product-type case is handled in Section~\ref{sec:general:product}.
The proof of Theorem~\ref{thm:main:symmetric} also spans two sections.
Once again, the affine case is handled in Section~\ref{sec:symmetric:affine} and the product-type case is handled in Section~\ref{sec:symmetric:product}.

%% file: 2.Preliminaries.tex
\section{Preliminaries} \label{sec:preliminaries}

\subsection{Problems and Definitions}

The framework of Holant problems is defined for functions mapping $[q]^k$ to $\mathbb{F}$ for a finite $q$ and some field $\mathbb{F}$.
In this paper, we investigate some of the tractable complex-weighted Boolean $\Holant$ problems, that is, all functions are of the type $[2]^k \to \mathbb{C}$.
Strictly speaking, for consideration of models of computation, functions take complex algebraic numbers.

A \emph{signature grid} $\Omega = (G, \mathcal{F}, \pi)$ consists of a graph $G = (V,E)$ and a set of constraint functions (also called signatures) $\mathcal{F}$,
where $\pi$ labels each vertex $v \in V$ and its incident edges with some $f_v \in \mathcal{F}$ and its input variables.
Note that in particular, $\pi$ specifies an ordering of edges/variables on each vertex.
The Holant problem on instance $\Omega$ is to evaluate $\Holant_\Omega = \sum_{\sigma} \prod_{v \in V} f_v(\sigma \mid_{E(v)})$,
a sum over all edge assignments $\sigma: E \to \{0,1\}$.

A function $f_v$ can be represented by listing its values in lexicographical order as in a truth table,
which is a vector in $\mathbb{C}^{2^{\deg(v)}}$.
Equivalently, $f_v$ can be treated as a tensor in $(\mathbb{C}^{2})^{\otimes \deg(v)}$.
We also use $f_{\textbf{x}}$ to denote the value $f(\textbf{x})$, where $\textbf{x}$ is a binary string.
A function $f \in \mathcal{F}$ is also called a \emph{signature}.
A symmetric signature $f$ on $k$ Boolean variables can be expressed as $[f_0,f_1,\dotsc,f_k]$,
where $f_w$ is the value of $f$ on inputs of Hamming weight $w$.

A Holant problem is parametrized by a set of signatures.
\begin{definition}
 Given a set of signatures $\mathcal{F}$,
 we define the counting problem $\Holant(\mathcal{F})$ as:

 Input: A \emph{signature grid} $\Omega = (G, \mathcal{F}, \pi)$;

 Output: $\Holant_\Omega$.
\end{definition}

A signature $f$ of arity $n$ is \emph{degenerate} if there exist unary signatures $u_j \in \mathbb{C}^2$ ($1 \le j \le n$)
such that $f = u_1 \otimes \cdots \otimes u_n$.
In a signature grid, it is equivalent to replace a degenerate one by corresponding unary signatures.
A symmetric degenerate signature has the form $u^{\otimes n}$, where the superscript denotes the tensor power.
Replacing a signature $f \in \mathcal{F}$ by a constant multiple $c f$,
where $c \ne 0$, does not change the complexity of $\Holant(\mathcal{F})$.
It introduces a global factor to $\Holant_\Omega$.

We say a signature set $\mathcal{F}$ is tractable (resp.~$\SHARPP$-hard)
if the corresponding counting problem $\Holant(\mathcal{F})$ can be solved in polynomial time (resp.~$\SHARPP$-hard).
Similarly for a signature $f$, we say $f$ is tractable (resp.~$\SHARPP$-hard) if $\{f\}$ is.

\subsection{Holographic Reduction}

To introduce the idea of holographic reductions, it is convenient to consider bipartite graphs.
We can always transform a general graph into a bipartite graph while preserving the Holant value, as follows.
For each edge in the graph, we replace it by a path of length two.
(This operation is called the \emph{2-stretch} of the graph and yields the edge-vertex incidence graph.)
Each new vertex is assigned the binary \textsc{Equality} signature $(=_2) = [1,0,1]$.

We use $\holant{\mathcal{F}}{\mathcal{G}}$ to denote the Holant problem on bipartite graphs $H = (U,V,E)$,
where each vertex in $U$ or $V$ is assigned a signature in $\mathcal{F}$ or $\mathcal{G}$, respectively.
An input instance for this bipartite Holant problem is a bipartite signature grid and is denoted by $\Omega = (H,\mathcal{F} \mid \mathcal{G},\pi)$.
Signatures in $\mathcal{F}$ are considered as row vectors (or covariant tensors);
signatures in $\mathcal{G}$ are considered as column vectors (or contravariant tensors)~\cite{DP91}.

For a 2-by-2 matrix $T$ and a signature set $\mathcal{F}$,
define $T \mathcal{F} = \{g \mid \exists f \in \mathcal{F}$ of arity $n,~g = T^{\otimes n} f\}$, similarly for $\mathcal{F} T$.
Whenever we write $T^{\otimes n} f$ or $T \mathcal{F}$,
we view the signatures as column vectors;
similarly for $f T^{\otimes n} $ or $\mathcal{F} T$ as row vectors.

Let $T$ be an element of $\mathbf{GL}_2(\mathbb{C})$, the group of invertible 2-by-2 complex matrices.
The holographic transformation defined by $T$ is the following operation:
given a signature grid $\Omega = (H,\mathcal{F} \mid \mathcal{G},\pi)$, for the same graph $H$,
we get a new grid $\Omega' = (H,\mathcal{F} T \mid T^{-1} \mathcal{G},\pi')$ by replacing 
$f\in\mathcal{F}$ (or $g\in \mathcal{G}$) with $T^{\otimes n}f$ (or $\left(T^{-1}\right)^{\otimes n} g$).

\begin{theorem}[Valiant's Holant Theorem~\cite{Val08}]
 If there is a holographic transformation mapping signature grid $\Omega$ to $\Omega'$,
 then $\Holant_\Omega = \Holant_{\Omega'}$.
\end{theorem}

Therefore, an invertible holographic transformation does not change the complexity of the Holant problem in the bipartite setting.
Furthermore, there is a particular kind of holographic transformation, the orthogonal transformation,
that preserves binary equality and thus can be used freely in the standard setting.
Let $\mathbf{O}_2(\mathbb{C})$ be the group of 2-by-2 complex matrices that are orthogonal.
Recall that a matrix $T$ is orthogonal if $T \transpose{T} = I$.

\begin{theorem}[Theorem~2.6 in~\cite{CLX11d}] \label{thm:orthogonal}
 Suppose $T \in \mathbf{O}_2(\mathbb{C})$ and let $\Omega = (H, \mathcal{F}, \pi)$ be a signature grid.
 Under a holographic transformation by $T$,
 we get a new grid $\Omega' = (H, T \mathcal F, \pi')$ and $\Holant_\Omega = \Holant_{\Omega'}$.
\end{theorem}

We also use $\mathbf{SO}_2(\mathbb{C})$ to denote the group of special orthogonal matrices,
i.e.~the subgroup of $\mathbf{O}_2(\mathbb{C})$ with determinant~1.

\subsection{Tractable Signature Sets without a Holographic Transformation}
\label{sec:prelim:tractable}

The following two signature sets are tractable without a holographic transformation~\cite{CLX14}.

\begin{definition}\label{def:affine}
 A $k$-ary function $f(x_1, \dotsc, x_k)$ is \emph{affine} if it has the form
 \[
  \lambda \cdot \chi_{A x = 0} \cdot i^{\sum_{j=1}^n \langle \mathbf{v}_j, x \rangle},
 \]
 where $\lambda \neq 0$ is in $\mathbb{C}$,
 $x = \transpose{(x_1, x_2, \dotsc, x_k, 1)}$,
 $A$ is a matrix over $\mathbb{F}_2$,
 $\mathbf{v}_j$ is a vector over $\mathbb{F}_2$ for each $j=1,\dots,n$,
 and $\chi$ is a 0-1 indicator function such that $\chi_{Ax = 0}$ is~$1$ iff $A x = 0$.
 Note that the dot product $\langle \mathbf{v}_j, x \rangle$ is calculated over $\mathbb{F}_2$,
 while the summation $\sum_{j=1}^n$ on the exponent of $i = \sqrt{-1}$ is evaluated as a sum mod~$4$ of 0-1 terms.
 We use $\mathscr{A}$ to denote the set of all affine functions.
\end{definition}

Notice that there is no restriction on the number of rows in the matrix $A$.
It is permissible that $A$ is the zero matrix so that $\chi_{A x = 0} = 1$ holds for all $x$.
An equivalent way to express the exponent of $i$ is as a quadratic polynomial (evaluated mod $4$) where all cross terms have an even coefficient.
This equivalent expression is often easier to use.

\begin{definition}
 A function is of \emph{product type} if it can be expressed as a function product of unary functions,
 binary equality functions $([1,0,1])$, and binary disequality functions $([0,1,0])$.
 We use $\mathscr{P}$ to denote the set of product-type functions.
\end{definition}

The above two types of functions, when restricted to be symmetric, have been characterized explicitly.
It has been shown (cf.~Lemma~2.2 in~\cite{HL16}) that if $f$ is a symmetric signature in $\mathscr{P}$,
then $f$ is either degenerate, binary disequality, or of the form $[a,0,\dotsc,0,b]$ for some $a, b \in \mathbb{C}$.
It is also known that (cf.~\cite{CLX11d}) the set of non-degenerate symmetric signatures in $\mathscr{A}$ is precisely the nonzero signatures
($\lambda \neq 0$) in $\mathscr{F}_1 \union \mathscr{F}_2 \union \mathscr{F}_3$\footnote{To be consistent with previous papers, we still use $\mathscr{F}_1$, $\mathscr{F}_2$, and $\mathscr{F}_3$ to denote the subclasses of $\mathscr{A}$. They are not to be confused with $\mathscr{A}_1$, $\mathscr{A}_2$, and $\mathscr{A}_3$ that will be introduced in Definition \ref{def:A123}.} 
with arity at least~$2$,
where $\mathscr{F}_1$,
$\mathscr{F}_2$,
and $\mathscr{F}_3$ are three families of signatures defined as
\begin{align*}
  \mathscr{F}_1 &= \left\{\lambda \left(\tbcolvec{1}{0}^{\otimes k} + i^r \tbcolvec{0}{1}^{\otimes k}\right) \st \lambda \in \mathbb{C}, k = 1, 2, \dotsc, r = 0, 1, 2, 3\right\},\\
 \mathscr{F}_2 &= \left\{\lambda \left(\tbcolvec{1}{1}^{\otimes k} + i^r \tbcolvec{1}{-1}^{\otimes k}\right) \st \lambda \in \mathbb{C}, k = 1, 2, \dotsc, r = 0, 1, 2, 3\right\}, \text{ and}\\
 \mathscr{F}_3 &= \left\{\lambda \left(\tbcolvec{1}{i}^{\otimes k} + i^r \tbcolvec{1}{-i}^{\otimes k}\right) \st \lambda \in \mathbb{C}, k = 1, 2, \dotsc, r = 0, 1, 2, 3\right\}.
\end{align*}
Let $\mathscr{F}_{123} = \mathscr{F}_1 \union \mathscr{F}_2 \union \mathscr{F}_3$ be the union of these three sets of signatures.
We explicitly list all the signatures in $\mathscr{F}_{123}$ (as row vectors) up to an arbitrary constant multiple from $\mathbb{C}$:\vspace{1em}
\\
\input{List_of_F1F2F3}

\subsection{\texorpdfstring{$\mathscr{A}$}{A}-transformable and \texorpdfstring{$\mathscr{P}$}{P}-transformable Signatures}

The tractable sets $\mathscr{A}$ and $\mathscr{P}$ are still tractable under a suitable holographic transformation.
This is captured by the following definition.

\begin{definition}
 A set $\mathcal{F}$ of signatures is $\mathscr{A}$-transformable (resp.~$\mathscr{P}$-transformable) if there exists
 a holographic transformation $T$ such that $\mathcal{F} \subseteq T \mathscr{A}$ (resp.~$\mathcal{F} \subseteq T \mathscr{P}$) 
 and $[1,0,1] T^{\otimes 2} \in \mathscr{A}$ (resp.~$[1,0,1] T^{\otimes 2} \in\mathscr{P}$).
\end{definition}

To refine the above definition,
we consider the stabilizer group of $\mathscr{A}$,
\[
 \StabA = \{T \in \mathbf{GL}_2(\mathbb{C}) \st T \mathscr{A} = \mathscr{A}\}.
\]
Technically what we defined is the left stabilizer group of $\mathscr{A}$,
but it turns out that the left and right stabilizer groups of $\mathscr{A}$ coincide \cite{CGW16}.

\begin{table}[htbp]
  \centering
  \begin{tabular}{|c|c|}
    \hline
    Name & Value \\
    \hline
    \hline
    $\alpha$ & $\sqrt{i} = e^{\frac{\pi i}{4}} = \frac{1+i}{\sqrt{2}}$ \\
    \hline
    $D$ & $\tbmatrix{1}{0}{0}{i}$ \\
    \hline
    $H_2$ & $\frac{1}{\sqrt{2}} \tbmatrix{1}{1}{1}{-1}$ \\
    \hline
    $X$ & $\tbmatrix{0}{1}{1}{0}$ \\
    \hline
    $Z$ & $\tfrac{1}{\sqrt{2}} \tbmatrix{1}{1}{i}{-i}$ \\
    \hline
  \end{tabular}
  \caption{Notations for some matrices and numbers}
  \label{tab:constants}
\end{table}

Some matrices and numbers are used extensively throughout the paper.
We summarize them in Table \ref{tab:constants}.
Note that $Z = D H_2$ and that $D^2  Z = \frac{1}{\sqrt{2}} \tbmatrix{1}{1}{-i}{i} = Z X$,
hence $X = Z^{-1} D^2  Z$.
It is easy to verify that $D,H_2,X,Z \in \StabA$.
In fact, $\StabA$ is precisely the set of nonzero scalar multiples of the group generated by $D$ and $H_2$ \cite{CGW16}.
Note that the zero matrix is not a stabilizer since $\mathscr{A}$ does not include the zero function.

The next lemma is the first step toward understanding $\mathscr{A}$-transformable signatures.
Recall that $\mathbf{O}_2(\mathbb{C})$ is the group of 2-by-2 orthogonal complex matrices.
The lemma shows that to determine $\mathscr{A}$-transformability, 
it is necessary and sufficient to consider only the orthogonal transformations and related ones.

\begin{lemma}[\cite{CGW16}] \label{lem:affine:trans}
 Let $\mathcal{F}$ be a set of signatures.
 Then $\mathcal{F}$ is $\mathscr{A}$-transformable iff
 there exists an $H \in \mathbf{O}_2(\mathbb{C})$ such that $\mathcal{F} \subseteq H \mathscr{A}$ 
 or $\mathcal{F} \subseteq H \tbmatrix{1}{0}{0}{\alpha} \mathscr{A}$.
\end{lemma}

Non-degenerate symmetric $\mathscr{A}$-transformable signatures are captured by three sets $\mathscr{A}_1$, $\mathscr{A}_2$, and $\mathscr{A}_3$,
which will be defined next
(not to be confused with $\mathscr{F}_1$, $\mathscr{F}_2$, and $\mathscr{F}_3$).

\begin{definition} \label{def:A123}
 A symmetric signature $f$ of arity $n$ is in, respectively, $\mathscr{A}_1$, or $\mathscr{A}_2$, or $\mathscr{A}_3$
 if there exists an $H \in \mathbf{O}_2(\mathbb{C})$ and a nonzero constant $c \in \mathbb{C}$ such that $f$ has the following form, respectively:
 \begin{itemize}
   \item $c H^{\otimes n} \left(\tbcolvec{1}{1}^{\otimes n}      + \beta \tbcolvec{1}{-1}^{\otimes n}\right)$, 
     where $\beta = \alpha^{tn+2r}$,  $r \in \{0,1,2,3\}$, and $t \in \{0,1\}$;
   \item or $c H^{\otimes n} \left(\tbcolvec{1}{i}^{\otimes n}      +       \tbcolvec{1}{-i}^{\otimes n}\right)$;
   \item or $c H^{\otimes n} \left(\tbcolvec{1}{\alpha}^{\otimes n} +   i^r \tbcolvec{1}{-\alpha}^{\otimes n}\right)$, where $r\in\{0,1,2,3\}$.
 \end{itemize} 
\end{definition}

For $i \in \{1,2,3\}$,
when such an orthogonal $H$ exists,
we say that $f \in \mathscr{A}_i$ with transformation $H$.
If $f \in \mathscr{A}_i$ with $I_2$, the identity matrix,
then we say $f$ is in the canonical form of $\mathscr{A}_i$.
Note that there is no direct correspondences between $(\mathscr{A}_i)$ and $(\mathscr{F}_i)$.

\begin{lemma}[\cite{CGW16}] \label{lem:cha:affine}
 Let $f$ be a non-degenerate symmetric signature.
 Then $f$ is $\mathscr{A}$-transformable iff $f \in \mathscr{A}_1 \union \mathscr{A}_2 \union \mathscr{A}_3$.
\end{lemma}

Analogous results hold for $\mathscr{P}$-transformable signatures.
Let the stabilizer group of $\mathscr{P}$ be
\[
 \StabP = \{T \in \mathbf{GL}_2(\mathbb{C}) \st T \mathscr{P} = \mathscr{P}\}.
\]
The group $\StabP$ is generated by (up to nonzero scalars) matrices of the form $\tbmatrix{1}{0}{0}{\nu}$ for any $\nu \in \mathbb{C}^*$ and $X = \tbmatrix{0}{1}{1}{0}$ \cite{CGW16}.

\begin{lemma}[\cite{CGW16}] \label{lem:product:trans}
 Let $\mathcal{F}$ be a set of signatures.
 Then $\mathcal{F}$ is $\mathscr{P}$-transformable
 iff there exists an $H \in \mathbf{O}_2(\mathbb{C})$ such that $\mathcal{F} \subseteq H \mathscr{P}$
 or $\mathcal{F} \subseteq H \tbmatrix{1}{1}{i}{-i} \mathscr{P}$.
\end{lemma}


\begin{definition}\label{def:p1}
 A symmetric signature $f$ of arity $n$ is in $\mathscr{P}_1$ if
 there exist an $H \in \mathbf{O}_2(\mathbb{C})$ and a nonzero $c \in \mathbb{C}$ such that
 $f = c H^{\otimes n} \left(\tbcolvec{1}{ 1}^{\otimes n}
                    + \beta \tbcolvec{1}{-1}^{\otimes n}\right)$,
 where $\beta \neq 0$.
\end{definition}

It is easy to check that $\mathscr{A}_1 \subset \mathscr{P}_1$.
We define $\mathscr{P}_2 = \mathscr{A}_2$.
For $i \in \{1,2\}$, when $H\in \mathbf{O}_2(\mathbb{C})$ exists (in Definition \ref{def:p1} and \ref{def:A123}, respectively), we say that $f \in \mathscr{P}_i$ with transformation $H$.
If $f \in \mathscr{P}_i$ with $I_2$,
then we say $f$ is in the canonical form of $\mathscr{P}_i$.

\begin{lemma}[\cite{CGW16}] \label{lem:cha:product}
 Let $f$ be a non-degenerate symmetric signature.
 Then $f$ is $\mathscr{P}$-transformable iff $f \in \mathscr{P}_1 \union \mathscr{P}_2$.
\end{lemma}

%% file: List_of_F1F2F3.tex
\parbox{0.61\textwidth}{
 \begin{enumerate}
  \item $[1, 0, \dotsc, 0, \pm 1]$; \hfill $(\mathscr{F}_1, r=0,2)$
  \item $[1, 0, \dotsc, 0, \pm i]$; \hfill $(\mathscr{F}_1, r=1,3)$
  \item $[1,  0, 1,  0, \dotsc,   0  \text{ or } 1]$; \hfill $(\mathscr{F}_2, r=0)$
  \item $[1, -i, 1, -i, \dotsc, (-i) \text{ or } 1]$; \hfill $(\mathscr{F}_2, r=1)$
  \item $[0,  1, 0,  1, \dotsc,   0  \text{ or } 1]$; \hfill $(\mathscr{F}_2, r=2)$
  \item $[1,  i, 1,  i, \dotsc,   i  \text{ or } 1]$; \hfill $(\mathscr{F}_2, r=3)$
  \item $[1,  0, -1,  0, 1,  0, -1,  0, \dotsc, 0 \text{ or } 1 \text{ or } (-1)]$; \hfill $(\mathscr{F}_3, r=0)$
  \item $[1,  1, -1, -1, 1,  1, -1, -1, \dotsc,               1 \text{ or } (-1)]$; \hfill $(\mathscr{F}_3, r=1)$
  \item $[0,  1,  0, -1, 0,  1,  0, -1, \dotsc, 0 \text{ or } 1 \text{ or } (-1)]$; \hfill $(\mathscr{F}_3, r=2)$
  \item $[1, -1, -1,  1, 1, -1, -1,  1, \dotsc,               1 \text{ or } (-1)]$. \hfill $(\mathscr{F}_3, r=3)$
 \end{enumerate}}

%% file: 3.Problems.tex
\def\problemSpace{4pt}

\section{Some Example Problems} \label{sec:problems}

In this section, we illustrate a few problems that are tractable via holographic reductions to affine or product-type functions.
Although the algorithms to solve them follow from a known paradigm, 
it is often non-trivial to find the correct holographic transformation.
Our main result provides a systematic way to search for these transformations.

\subsection{A Fibonacci-like Problem}

Fibonacci gates were introduced in~\cite{CLX11c}.
They define tractable counting problems,
and holographic algorithms based on Fibonacci gates work over general (i.e.~not necessarily planar) graphs.
However, Fibonacci gates are symmetric by definition.
An example of a Fibonacci gate is the signature $f = [f_0,f_1,f_2,f_3] = [1,0,1,1]$.
Its entries satisfy the recurrence relation of the Fibonacci numbers, i.e.~$f_2 = f_1 + f_0$ and $f_3 = f_2 + f_1$.
For $\Holant(f)$, the input is a 3-regular graph,
and the problem is to count spanning subgraphs such that no vertex has degree~1.

A symmetric signature $g = [g_0, g_1, \dotsc, g_n]$ is a generalized Fibonacci gate of type $\lambda \in \mathbb{C}$
if $g_{k+2} = \lambda g_{k+1} + g_k$ holds for all $k \in \{0,1,\dotsc,n-2\}$.
The standard Fibonacci gates are of type $\lambda = 1$.
An example of a generalized Fibonacci gate is $g = [3,1,3,1]$, which has type $\lambda = 0$.
In contrast to $\Holant(f)$, the problem $\Holant(g)$ permits all possible spanning subgraphs.
The output is the sum of the weights of each spanning subgraph.
The weight of a spanning subgraph $S$ is $3^{k(S)}$,
where $k(S)$ is the number of vertices of even degree in $S$.
Since $g = [3,1,3,1]$ is Fibonacci, the problem $\Holant(g)$ is computable in polynomial time \cite{CLX11d,CGW16}.
One new family of holographic algorithms in this paper extends Fibonacci gates to asymmetric signatures.

In full notation, the ternary signature $g$ is $\transpose{(3,1,1,3,1,3,3,1)}$.
Consider the asymmetric signature $h = \transpose{(3,1,-1,-3,-1,-3,3,1)}$.
This signature $h$ differs from $g$ by a negative sign in four entries.
Although $h$ is not a generalized Fibonacci gate or even a symmetric signature,
it still defines a tractable Holant problem.
Under a holographic transformation by $Z^{-1}$, where $Z = \frac{1}{\sqrt{2}} \tbmatrix{1}{1}{i}{-i}$,
\begin{align*}
 \Holant(h)
 = \holant{{=}_2}{h}
 = \holant{{=}_2 (Z^{-1})^{\otimes 2}}{Z^{\otimes 3} h}
 = \holant{[1,0,-1]}{\hat{h}},
\end{align*}
where $\hat{h} = 2 i \sqrt{2} (0,1,0,0,0,0,2i,0)$.
Both $[1,0,-1](x_1,x_2) = \textsc{Equality}(x_1,x_2) \cdot [1,-1](x_1)$ and
$\hat{h}(x_1,x_2,x_3) = 2 i \sqrt{2} \cdot \textsc{Equality}(x_1,x_2) \cdot \textsc{Disequality}(x_2,x_3) \cdot [1,2i](x_1)$ are product-type signatures.

It turns out that for all values of $\lambda \neq \pm 2 i$,
the generalized Fibonacci gates of type $\lambda$ are $\mathscr{P}$-transformable.
The value of $\lambda$ indicates under which holographic transformation the signatures become product type.
For $\lambda = \pm 2 i$, the generalized Fibonacci gates of type $\lambda$ are vanishing,
which means the output is always zero for every possible input (see~\cite{CGW16} for more on vanishing signatures).

\subsection{Some Cycle Cover Problems and Orientation Problems}

To express some problems involving asymmetric signatures of arity~4,
it is convenient to arrange the~16 outputs into a 4-by-4 matrix.
With a slight abuse of notation, we also write a function $f(x_1, x_2, x_3, x_4)$ in its matrix form, namely $f =
  \left[
  \begin{smallmatrix}
   f_{0000} & f_{0010} & f_{0001} & f_{0011}\\
   f_{0100} & f_{0110} & f_{0101} & f_{0111}\\
   f_{1000} & f_{1010} & f_{1001} & f_{1011}\\
   f_{1100} & f_{1110} & f_{1101} & f_{1111}
  \end{smallmatrix}
  \right]$,
 where the row is indexed by two bits $(x_1,x_2)$ and the column is
indexed by two bits $(x_4,x_3)$ in reverse order.
We call this the \emph{signature matrix}.


Consider the problem of counting the number of cycle covers in a given graph.
This problem is $\SHARPP$-hard even when restricted to planar 4-regular graphs~\cite{GW13}.
As a Holant problem, its expression is $\Holant(f)$, where $f(x_1, x_2, x_3, x_4)$ is the symmetric signature $[0,0,1,0,0]$.
The signature matrix of $f$ is
 $
 \left[
 \begin{smallmatrix}
  0 & 0 & 0 & 1\\
  0 & 1 & 1 & 0\\
  0 & 1 & 1 & 0\\
  1 & 0 & 0 & 0
 \end{smallmatrix}
 \right]$.
The six entries in the support of $f$, which are all of Hamming weight two
(indicating that a cycle cover passes through each vertex exactly twice),
can be divided into two parts, namely $\{0011, 0110, 1100, 1001\}$ and $\{0101, 1010\}$.
In the planar setting, this corresponds to a pairing of consecutive or non-consecutive incident edges.
Both sets are invariant under cyclic permutations.

Suppose we removed the inputs $0101$ and $1010$ from the support of $f$,
which are the two 1's on the anti-diagonal in the middle of $M_f$.
Call the resulting signature $g$, which has signature matrix
 $
 \left[
 \begin{smallmatrix}
  0 & 0 & 0 & 1\\
  0 & 1 & 0 & 0\\
  0 & 0 & 1 & 0\\
  1 & 0 & 0 & 0
 \end{smallmatrix}
 \right]$.\footnote{Recall that in general we require the input signature grid to specify the ordering of the edges (namely variables) on each vertex.
 This is not necessary for symmetric signatures, but when asymmetric signatures are involved, specifying the ordering is essential.}
These new 0's impose a constraint on the types of cycle covers allowed.
We call a cycle cover \emph{valid} if it satisfies this new constraint.
A valid cycle cover must not pass through a vertex in a ``crossing'' way.
Counting the number of such cycle covers over 4-regular graphs can be done in polynomial time,
even without the planarity restriction.
The signature $g(x_1, x_2, x_3, x_4) = \textsc{Dis-Equality} (x_1, x_3) \cdot \textsc{Dis-Equality}(x_2, x_4)$ is of the product type $\mathscr{P}$,
therefore $\Holant(g)$ is tractable.

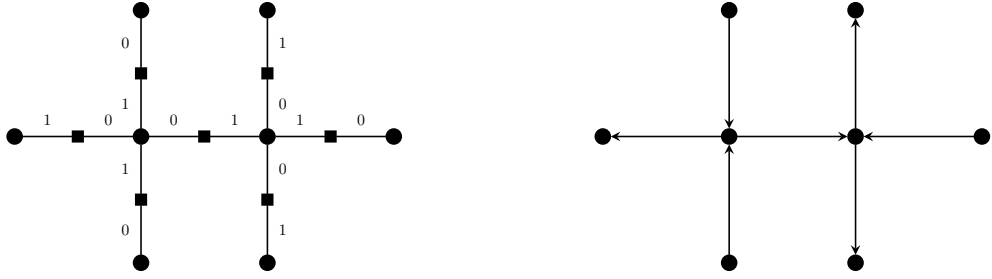
\begin{figure}[t]
 \centering
 \def\capWidth{7cm}
 \def\arrowType{stealth}
 \captionsetup[subfigure]{width=\capWidth}
 \subfloat[An admissible assignment to this graph fragment. The circle vertices are assigned $\hat{g}$ and the square vertices are assigned $\neq_2$.]{
  \makebox[\capWidth][c]{
   \begin{tikzpicture}[scale=\scale,transform shape,node distance=\nodeDist,semithick]
    \node[square]    (0)               {};
    \node[internal]  (1) [left  of=0]  {};
    \node[internal]  (2) [right of=0]  {};
    \node[square]    (3) [above of=1]  {};
    \node[square]    (4) [above of=2]  {};
    \node[internal]  (5) [above of=3]  {};
    \node[internal]  (6) [above of=4]  {};
    \node[square]    (7) [left  of=1]  {};
    \node[square]    (8) [right of=2]  {};
    \node[internal]  (9) [left  of=7]  {};
    \node[internal] (10) [right of=8]  {};
    \node[square]   (11) [below of=1]  {};
    \node[square]   (12) [below of=2]  {};
    \node[internal] (13) [below of=11] {};
    \node[internal] (14) [below of=12] {};
    \path (1) edge node[label=above:0] {}  (0)
              edge node[label=left :1] {}  (3)
              edge node[label=above:0] {}  (7)
              edge node[label=left :1] {} (11)
          (2) edge node[label=above:1] {}  (0)
              edge node[label=right:0] {}  (4)
              edge node[label=above:1] {}  (8)
              edge node[label=right:0] {} (12)
          (5) edge node[label=left :0] {}  (3)
          (6) edge node[label=right:1] {}  (4)
          (9) edge node[label=above:1] {}  (7)
         (10) edge node[label=above:0] {}  (8)
         (13) edge node[label=left :0] {} (11)
         (14) edge node[label=right:1] {} (12);
   \end{tikzpicture} \label{subfig:disequality_assignment}}}
 \qquad
 \subfloat[The orientation induced by the assignment in~\protect\subref{subfig:disequality_assignment}.]{
  \makebox[\capWidth][c]{
   \begin{tikzpicture}[scale=\scale,transform shape,>=\arrowType, node distance=\nodeDist,semithick]
    \node[external]  (0)               {};
    \node[internal]  (1) [left  of=0]  {};
    \node[internal]  (2) [right of=0]  {};
    \node[external]  (3) [above of=1]  {};
    \node[external]  (4) [above of=2]  {};
    \node[internal]  (5) [above of=3]  {};
    \node[internal]  (6) [above of=4]  {};
    \node[external]  (7) [left  of=1]  {};
    \node[external]  (8) [right of=2]  {};
    \node[internal]  (9) [left  of=7]  {};
    \node[internal] (10) [right of=8]  {};
    \node[external] (11) [below of=1]  {};
    \node[external] (12) [below of=2]  {};
    \node[internal] (13) [below of=11] {};
    \node[internal] (14) [below of=12] {};
    \path (1) edge[->]  (2)
              edge[<-]  (5)
              edge[->]  (9)
              edge[<-] (13)
          (2) edge[->]  (6)
              edge[<-] (10)
              edge[->] (14);
   \end{tikzpicture} \label{subfig:orientation}}}
 \caption{A fragment of an instance to $\holant{{\ne}_2}{\hat{g}}$, which must be a $(2,4)$-regular bipartite graph.
          Note the saddle orientation of the edges incident to the two vertices with all four edges depicted.}
 \label{fig:instance_fragment}
\end{figure}

Under a holographic transformation by $Z = \frac{1}{\sqrt{2}} \tbmatrix{1}{1}{i}{-i}$,
we obtain the problem
\begin{align*}
 \Holant(g)
 = \holant{{=}_2}{g}
 = \holant{{=}_2 Z^{\otimes 2}}{(Z^{-1})^{\otimes 4} g}
 = \holant{{\ne}_2}{\hat{g}},
\end{align*}
where
$\hat{g}
:=(Z^{-1})^{\otimes 4} g = 
\left[
\begin{smallmatrix}
 -1 & 0 & 0 & 0\\
  0 & 0 & 1 & 0\\
  0 & 1 & 0 & 0\\
  0 & 0 & 0 & -1
\end{smallmatrix}
\right]$.
This problem has the following interpretation.
It is a Holant problem on bipartite graphs.
On the right side of the bipartite graph, the vertices must all have degree~4 and are assigned the signature $\hat{g}$.
On the left side, the vertices must all have degree~2 and are assigned the binary disequality constraint $\ne_2$.
The disequality constraints suggest an orientation between their two neighboring vertices of degree~4 (see Figure~\ref{fig:instance_fragment}).
By convention, we view the edge as having its tail assigned~0 and its head assigned~1.
Then every valid assignment in this bipartite graph naturally corresponds to an orientation in the original 4-regular graph.

If the four inputs $0011$, $0110$, $1100$, and $1001$ were in the support of $\hat{g}$,
then the Holant sum would be over all possible orientations with an even number of incoming edges at each vertex.
As it is, the sum is over all possible orientations with an even number of incoming edges at each vertex that also 
forbid those four types of orientations at each vertex,
as specified by $\hat{g}$.
The following orientations are admissible by $\hat{g}$:
The orientation of the edges are such that
at each vertex all edges are oriented out (source vertex),
or all edges are oriented in (sink vertex),
or the edges are cyclically oriented in, out, in, out (saddle vertex).

Thus, the output of $\holant{{\ne}_2}{\hat{g}}$ is a weighted sum over of these admissible orientations.
Each admissible orientation $O$ contributes a weight $(-1)^{s(O)}$ to the sum,
where $s(O)$ is the number of source and sink vertices in an orientation $O$.
We can express this as $\sum_{O \in \mathcal{O}(G)} (-1)^{s(O)}$,
where $\mathcal{O}(G)$ is the set of admissible orientations for $G$,
which are those orientations that only contain source, sink, and saddle vertices.
In words, the value is the number of admissible orientations with an even number of sources and sinks
minus the number of admissible orientations with an odd number of sources and sinks.
This orientation problem may seem quite different from the restricted cycle cover problem we started with,
but they are, in fact, the same problem.
Since $\Holant(g)$ is tractable, so is $\holant{{\ne}_2}{\hat{g}}$.

Now, consider a slight generalization of this orientation problem.

\vspace{\problemSpace}
\textbf{Problem:} \#$\lambda$\textsc{-SourceSinkSaddleOrientations}

\textbf{Input:} An undirected 4-regular graph $G$ (equipped with a local edge-ordering on every vertex).

\textbf{Output:} $\sum_{O \in \mathcal{O}(G)} \lambda^{s(O)}$.
\vspace{\problemSpace}

\noindent
For $\lambda = -1$, we recover the orientation problem from above.
For $\lambda = 1$, the problem is also tractable since, when viewed as a bipartite Holant problem on the $(2,4)$-regular bipartite vertex-edge incidence graph,
the disequality constraint on the vertices of degree~2 and the constraint on the vertices of degree~4 are both product-type functions.
As a function of $x_1,x_2,x_3,x_4$, the constraint on the degree~4 vertices is $\textsc{Equality}(x_1,x_3) \cdot \textsc{Equality}(x_2,x_4)$.
Let $s_{k,m}(G)$ be the number of $O \in \mathcal{O}(G)$ such that $s(O) \equiv k \pmod{m}$.
Then the output of this problem with $\lambda = 1$ is $s_{0,2}(G) + s_{1,2}(G)$ and the output of this problem with $\lambda = -1$ is $s_{0,2}(G) - s_{1,2}(G)$.
Therefore, we can compute both $s_{0,2}(G)$ and $s_{1,2}(G)$.
However, more is possible.

For $\lambda = i$, the problem is tractable using affine constraints.
In the $(2,4)$-regular bipartite vertex-edge incidence graph,
the disequality constraint assigned to the vertices of degree~2 is affine.
On the vertices of degree~4, the assigned constraint function is an affine signature since
the affine support is defined by the affine linear system $x_1 = x_3$ and $x_2 = x_4$
while the quadratic polynomial in the exponent of $i$ is $2 x_1 x_2 + 3 x_1 + 3 x_2 + 1$.
(Recall that in the definition of $\mathscr{A}$, Definition~\ref{def:affine}, we need to evaluate the quadratic polynomial mod $4$ instead of $2$,
and $x^2=x$ for any $x\in\{0,1\}$.)
Although the output is a complex number,
the real and imaginary parts encode separate information.
The real part is $s_{0,4}(G) - s_{2,4}(G)$ and the imaginary part is $s_{1,4}(G) - s_{3,4}(G)$.
Since $s_{0,2}(G) = s_{0,4}(G) + s_{2,4}(G)$ and $s_{1,2}(G) = s_{1,4}(G) + s_{3,4}(G)$,
we can actually compute all four quantities $s_{0,4}(G)$, $s_{1,4}(G)$, $s_{2,4}(G)$, and $s_{3,4}(G)$ in polynomial time.

\subsection{An Enigmatic Problem}

Some problems may be a challenge for the human intelligence to grasp. 
But in a platonic view of computational complexity, they are no less valid problems.

For example, consider the problem $\Holant((1+c^2)^{-1} [1,0,-i] \mid f)$ where $f$ has the signature matrix
\tiny
\[
  \left[
  \begin{smallmatrix}
   0
   &  (4 + 4 i) \left(28 + 20 \sqrt{2} +   \sqrt{  2 \left(799 + 565 \sqrt{2}\right)}\right)
   &  (4 + 4 i) \left(28 + 20 \sqrt{2} +   \sqrt{  2 \left(799 + 565 \sqrt{2}\right)}\right)
   &      -8 i  \left(13 +  9 \sqrt{2} + 2 \sqrt{ 82 + 58 \sqrt{2}}\right)
   \\
      (4 + 4 i) \left(28 + 20 \sqrt{2} +   \sqrt{  2 \left(799 + 565 \sqrt{2}\right)}\right)
   &      -8 i  \left(13 +  9 \sqrt{2} + 2 \sqrt{ 82 + 58 \sqrt{2}}\right)
   &       8 i  \left(18 + 13 \sqrt{2} + 4 \sqrt{ 41 + 29 \sqrt{2}}\right)
   & (-4 + 4 i) \left(12 +  8 \sqrt{2} +   \sqrt{274 + 194 \sqrt{2}}\right)
   \\
      (4 + 4 i) \left(28 + 20 \sqrt{2} +   \sqrt{  2 \left(799 + 565 \sqrt{2}\right)}\right)
   &       8 i  \left(18 + 13 \sqrt{2} + 4 \sqrt{ 41 +  29 \sqrt{2}}\right)
   &      -8 i  \left(13 +  9 \sqrt{2} + 2 \sqrt{ 82 +  58 \sqrt{2}}\right)
   & (-4 + 4 i) \left(12 +  8 \sqrt{2} +   \sqrt{274 + 194 \sqrt{2}}\right)
   \\
          -8 i  \left(13 +  9 \sqrt{2} + 2 \sqrt{ 82 +  58 \sqrt{2}}\right)
   & (-4 + 4 i) \left(12 +  8 \sqrt{2} +   \sqrt{274 + 194 \sqrt{2}}\right)
   & (-4 + 4 i) \left(12 +  8 \sqrt{2} +   \sqrt{274 + 194 \sqrt{2}}\right)
   & -16        \left(13 +  9 \sqrt{2} + 2 \sqrt{ 82 +  58 \sqrt{2}}\right)
  \end{smallmatrix}
  \right]
\]
\normalsize
and $c = 1 + \sqrt{2} + \sqrt{2 (1 + \sqrt{2})}$.
Most likely no one has ever considered this problem before.
Yet this nameless problem is $\mathscr{A}$-transformable under $T = \tbmatrix{1}{0}{0}{\alpha} \tbmatrix{1}{c}{-c}{1}$,
and hence it is really the \emph{same} problem as a more comprehensible problem defined by $\hat{f}=(T^{-1})^{\otimes 4} f$.
Namely,
\small
\[
  \Holant((1+c^2)^{-1} [1,0,-i] \: | \: f)
  = \Holant((1+c^2)^{-1} [1,0,-i] T^{\otimes 2} \: | \: (T^{-1})^{\otimes 4} f)
  = \Holant([1,0,1] \: | \: \hat{f})
  = \Holant(\hat{f}),
\]
\normalsize
where $
 \hat{f}
 =
 \left[
 \begin{smallmatrix}
  \phantom{-}1 &           -1 &           -1 &           -1\\
            -1 &           -1 & \phantom{-}1 &           -1\\
            -1 & \phantom{-}1 &           -1 &           -1\\
            -1 &           -1 &           -1 & \phantom{-}1
 \end{smallmatrix}
 \right]$.
We can express $\hat{f}$ as $\hat{f}(x_1,x_2,x_3,x_4) = i^{Q(x)}$,
where $Q(x_1,x_2,x_3,x_4) = 2 (x_1^2 + x_2^2 + x_3^2 + x_4^2 + x_1 x_2 + x_2 x_3 + x_3 x_4 + x_4 x_1)$.
Therefore, $\hat{f}$ is affine, which means that $\Holant(\hat{f})$ as well as $\Holant((1+c^2)^{-1} [1,0,-i] \mid f)$ are tractable.
Furthermore, notice that $\hat{f}$ only contains integers even though $(1 + c^2)^{-1} [1,0,-i]$ and $f$ contain many complex numbers with irrational real and imaginary parts.
Thus, $\Holant((1 + c^2)^{-1} [1,0,-i] \mid f)$ is not only tractable, but it always outputs an integer.
Apparent anomalies like $\Holant((1 + c^2)^{-1} [1,0,-i] \mid f)$, however contrived they may seem to be to the human eye,
behoove the creation of a systematic theory to understand and characterize the tractable cases.

%% file: 4.General.Affine.tex
\section{General \texorpdfstring{$\mathscr{A}$}{A}-transformable Signatures} \label{sec:general:affine}

In this section, we give the algorithm to check $\mathscr{A}$-transformable signatures.
Our general strategy is to bound the number of possible transformations by a polynomial in the length of the function,
and then enumerate all of them.
There are some cases where this number cannot be bounded, and those cases are handled separately.

Let $f$ be a signature of arity $n$.
It is given as a column vector in $\mathbb{C}^{2^n}$ with bit length $N$,
which is on the order of $2^n$.
We denote its entries by $f_{\mathbf{x}} = f(\mathbf{x})$ indexed by $\mathbf{x}\in\{0,1\}^n$.
The entries are from a fixed degree algebraic extension of $\mathbb{Q}$ and we may assume basic bit operations in the field take unit time.

Notice that the number of general affine signatures of arity $n$ is on the order of $2^{n^2}$.
Hence a naive check of the membership of affine signatures would result in a super-polynomial running time in $N$.
Instead, we present a polynomial-time algorithm.

\begin{lemma} \label{lem:general:affine:test}
 There is an algorithm to decide whether a given signature $f$ of arity $n$ belongs to $\mathscr{A}$ with running time polynomial in $N$,
 the bit length of $f$.
\end{lemma}

\begin{proof}
 We may assume that $f$ is not identically zero.
 Normalize $f$ so that the first nonzero entry of $f$ is $1$.
 If there exists a nonzero entry of $f$ after normalization that is not a power of $i$,
 then $f \not \in \mathscr{A}$, so assume that all entries are now powers of $i$.

 The next step is to decide if the support $\mathcal{S} \neq \emptyset$ of $f$ forms an affine linear subspace.
 We try to build a basis for $\mathcal{S}$ inductively.
 It may end successfully or find an inconsistency.
 We choose the index of the first nonzero entry $\mathbf{b}_0 \in \mathcal{S}$ as our first basis element.
 Assume we have a set of basis elements $\mathcal{B} = \{\mathbf{b}_0, \ldots, \mathbf{b}_k\} \subseteq \mathcal{S}$.
 Consider the affine linear span $\operatorname{Span}(\mathcal{B})$.
 We check if $\operatorname{Span}(\mathcal{B}) \subseteq \mathcal{S}$.
 If not, then $\mathcal{S}$ is not affine and $f \not\in \mathscr{A}$, so suppose that this is the case.
 If $\operatorname{Span}(\mathcal{B}) = \mathcal{S}$, then we are done.
 Lastly, if $\mathcal{S} - \operatorname{Span}(\mathcal{B}) \neq \emptyset$,
 then pick the next element $\mathbf{b}_{k+1} \in \mathcal{S} - \operatorname{Span}(\mathcal{B})$.
 Let $\mathcal{B}' = \mathcal{B} \union \{\mathbf{b}_{k+1}\}$ and repeat with the new basis set $\mathcal{B}'$.

 Now assume that $\mathcal{S}$ is an affine subspace,
 that we have a linear system defining it,
 and that every nonzero entry of $f$ is a power of $i$.
 If $\mathcal{S}$ has dimension 0, then $\mathcal{S}$ is a single point, and $f \in \mathscr{A}$.
 Otherwise, $\operatorname{dim}(\mathcal{S}) = r \ge 1$, and (after reordering) $x_1, \dots, x_r$ are free variables of the linear system defining $\mathcal{S}$.
 For each $\mathbf{x} \in \{0,1\}^r$,
 let $\mathbf{y} \in \{0,1\}^{n-r}$ be the unique extension such that $\mathbf{x} \mathbf{y} \in \mathcal{S}$.
 For each $\mathbf{x}$, define $p_\mathbf{x} \in \mathbb{Z}_4$ such that $f_{\mathbf{x} \mathbf{y}} = i^{p_\mathbf{x}} \neq 0$.
 We will use the alternative expression for affine functions: namely, we want to decide if there exists a quadratic polynomial
 \begin{align*}
  Q(\mathbf{x}) = \sum_{j=1}^r c_j x_j^2 + 2 \sum_{1 \leq k < \ell \leq r} c_{k \ell} x_k x_\ell + c,
 \end{align*}
 where $c, c_j, c_{k \ell} \in \mathbb{Z}_4$,
 for $1 \leq j \leq r$ and $1 \leq k < \ell \leq r$,
 such that $Q(\mathbf{x}) \equiv p_\mathbf{x} \pmod{4}$ for all $\mathbf{x} \in \{0,1\}^r$.
 Setting $\mathbf{x} = \mathbf{0} \in\{0,1\}^r$ determines $c$.
 Setting exactly one $x_j = 1$ and the rest to 0 determines $c_j$.
 Setting exactly two $x_k = x_\ell = 1$ and the rest to 0 determines $c_{k \ell}$.
 Then we verify if $Q(\mathbf{x})$ is consistent with $f$, and $f \in \mathscr{A}$ iff it is so.
\end{proof}

For later use, we note the following corollary. 

\begin{corollary} \label{cor:general:affine:test:alpha}
 There is an algorithm to decide whether a given signature $f$ of arity $n$ belongs to $\tbmatrix{1}{0}{0}{\alpha} \mathscr{A}$ with running time polynomial in $N$,
 the bit length of $f$.
\end{corollary}

\begin{proof}
 For $\arity(f) = n$, just check if $\tbmatrix{1}{0}{0}{\alpha^{-1}}^{\otimes n} f \in \mathscr{A}$ by Lemma~\ref{lem:general:affine:test}.
\end{proof}

We can strengthen Lemma~\ref{lem:affine:trans} by restricting to orthogonal transformations within $\mathbf{SO}_2(\mathbb{C})$.

\begin{lemma} \label{lem:affine:trans:so2}
 Let $\mathcal{F}$ be a set of signatures.
 Then $\mathcal{F}$ is $\mathscr{A}$-transformable iff
 there exists an $H \in \mathbf{SO}_2(\mathbb{C})$ such that $\mathcal{F} \subseteq H \mathscr{A}$ 
 or $\mathcal{F} \subseteq H \tbmatrix{1}{0}{0}{\alpha} \mathscr{A}$.
\end{lemma}

\begin{proof}
 Sufficiency is obvious by Lemma~\ref{lem:affine:trans}.

 Assume that $\mathcal{F}$ is $\mathscr{A}$-transformable.
 By Lemma~\ref{lem:affine:trans}, there exists an $H \in \mathbf{O}_2(\mathbb{C})$
 such that $\mathcal{F} \subseteq H \mathscr{A}$ or $\mathcal{F} \subseteq H \tbmatrix{1}{0}{0}{\alpha} \mathscr{A}$.
 If $H \in\mathbf{SO}_2(\mathbb{C})$, we are done,
 so assume that $H \in \mathbf{O}_2(\mathbb{C}) \setminus \mathbf{SO}_2(\mathbb{C})$.
 We want to find an $H' \in \mathbf{SO}_2(\mathbb{C})$ such that
 $\mathcal{F} \subseteq H' \mathscr{A}$ or $\mathcal{F}\subseteq H' \tbmatrix{1}{0}{0}{\alpha} \mathscr{A}$.
 Let $H' = H \tbmatrix{1}{0}{0}{-1} \in \mathbf{SO}_2(\mathbb{C})$.
 There are two cases to consider.
 \begin{enumerate}
  \item Suppose $\mathcal{F} \subseteq H \mathscr{A}$.
  Then since $\tbmatrix{1}{0}{0}{-1} \in \StabA$,
  \begin{align*}
   \mathcal{F}
   &\subseteq H \tbmatrix{1}{0}{0}{-1} \mathscr{A}\\
   &= H' \mathscr{A}.
  \end{align*}
  \item Suppose $\mathcal{F} \subseteq H \tbmatrix{1}{0}{0}{\alpha} \mathscr{A}$.
  Then since $\tbmatrix{1}{0}{0}{-1} \in \StabA$ commutes with $\tbmatrix{1}{0}{0}{\alpha}$,
  \begin{align*}
   \mathcal{F}
   &\subseteq H \tbmatrix{1}{0}{0}{\alpha} \tbmatrix{1}{0}{0}{-1} \mathscr{A}\\
   &= H \tbmatrix{1}{0}{0}{-1} \tbmatrix{1}{0}{0}{\alpha} \mathscr{A}\\
   &= H' \tbmatrix{1}{0}{0}{\alpha} \mathscr{A}.
   \qedhere
  \end{align*}
 \end{enumerate}
\end{proof}

We now observe some properties of a signature under transformations in $\mathbf{SO}_2(\mathbb{C})$.
Let $f$ be a signature and $H = \tbmatrix{a}{b}{-b}{a} \in \mathbf{SO}_2(\mathbb{C})$ where $a^2 + b^2 = 1$.
Notice that $v_0 = (1,i)$ and $v_1 = (1,-i)$ are row eigenvectors of $H$ with eigenvalues $a - b i$ and $a + b i$ respectively.
Let $Z' = \tbmatrix{1}{i}{1}{-i}$. Then $Z' H = T Z'$,
where $T = \tbmatrix{a-bi}{0}{0}{a+bi}$.

For an index or a bit-string $\mathbf{u} = (u_1, \ldots, u_n) \in \{0,1\}^n$ of length $n$, let
\[
 v_{\mathbf{u}} := v_{u_1} \otimes v_{u_2} \otimes \ldots \otimes v_{u_n},
\]
and let $wt(\mathbf{u})$ be the Hamming weight of $\mathbf{u}$.
Then $v_\mathbf{u}$ is a row eigenvector of the $2^n$-by-$2^n$ matrix $H^{\otimes n}$ with eigenvalue 
\begin{align}\label{eqn:eiganvalue}
  (a - b i)^{n - wt(\mathbf{u})} (a + b i)^{wt(\mathbf{u})}
= (a - b i)^{n - 2 wt(\mathbf{u})}
= (a + b i)^{2 wt(\mathbf{u}) - n}
\end{align}
since $(a + b i) (a - b i) = a^2 + b^2 = 1$.
In this paper, the following $Z'$-transformation plays an important role.
For any function $f$ on $\{0, 1\}^n$, we define
\[
 \hat{f} = Z'^{\otimes n} f.
\]
Then $\hat{f}_\mathbf{u}= \langle v_\mathbf{u}, f \rangle$,
as a dot product.

\begin{lemma} \label{lem:ortho:diag}
 Suppose $f$ and $g$ are signatures of arity $n$ and let $H = \tbmatrix{a}{b}{-b}{a}$ and $T = \tbmatrix{a-bi}{0}{0}{a+bi}$.
 Then $g = H^{\otimes n} f$ iff $\hat{g} = T^{\otimes n} \hat{f}$.
\end{lemma}

\begin{proof}
 Since $Z' H = T Z'$,
 \begin{align*}
  g = H^{\otimes n} f
  &\iff Z'^{\otimes n} g = Z'^{\otimes n} H^{\otimes n} f\\
  &\iff Z'^{\otimes n} g = T^{\otimes n} Z'^{\otimes n} f\\
  &\iff \phantom{Z'^{\otimes n}} \hat{g} = T^{\otimes n} \hat{f}.
  \qedhere
 \end{align*}
\end{proof}

We note that $\transpose{v_\mathbf{u}}$ is also a column eigenvector of $H^{\otimes n}$ with eigenvalue $(a - b i)^{2 wt(\mathbf{u}) - n}$.
Now we characterize the signatures that are invariant under transformations in $\mathbf{SO}_2(\mathbb{C})$.

\begin{lemma} \label{lem:ortho:invariant}
 Let $f$ be a signature.
 Then $f$ is invariant under transformations in $\mathbf{SO}_2(\mathbb{C})$ (up to a nonzero constant)
 iff the support of $\hat{f}$ contains at most one Hamming weight.
\end{lemma}

\begin{proof}
 This clearly holds when $f$ is identically zero, so assume that $f$ contains a nonzero entry and has arity $n$.
 Such an $f$ is invariant under any $H$ (up to a nonzero constant) iff $f$ is a column eigenvector of $H^{\otimes n}$.
 Consider $H= \tbmatrix{a}{b}{-b}{a} \in \mathbf{SO}_2(\mathbb{C})$ where $a^2+b^2=1$.
 Then $H^{\otimes n}$ has $n+1$ distinct eigenvalues $(a-bi)^{n - w} (a+bi)^{w}$, for $0 \le w \le n$.
 As a consequence, $f$ is a column eigenvector of $H^{\otimes n}$ iff $f$ is a nonzero linear combination of $\transpose{v_{\mathbf{u}}}$ of the same Hamming weight $wt(\mathbf{u})$.
 Hence $f$ is invariant under $H$ iff the support of $\hat{f}$ contains at most one Hamming weight.
\end{proof}

Using Lemma~\ref{lem:ortho:invariant},
we can efficiently decide if there exists an $H \in \mathbf{SO}_2(\mathbb{C})$ such that $H^{\otimes n} f \in \mathscr{A}$.

\begin{lemma} \label{lem:general:affine:test:ortho}
 There is an algorithm to decide in time polynomial in $N$, for any input signature $f$ of arity $n$,
 whether there exists an $H \in \mathbf{SO}_2(\mathbb{C})$ such that $H^{\otimes n}f \in \mathscr{A}$.
 If so, either $f\in\mathscr{A}$ and $f$ is invariant under any transformation in $\mathbf{SO}_2(\mathbb{C})$,
 or there exist at most $8n$ many $H \in \mathbf{SO}_2(\mathbb{C})$ such that $H^{\otimes n} f\in\mathscr{A}$,
 and they can all be computed in time polynomial in $N$.
\end{lemma}

\begin{proof}
 Compute $\hat{f} = Z'^{\otimes n} f$.
 If the support of $\hat{f}$ contains at most one Hamming weight,
 then by Lemma~\ref{lem:ortho:invariant},
 $f$ is invariant under any $H \in \mathbf{SO}_2(\mathbb{C})$.
 Therefore we only need to directly decide if $f \in \mathscr{A}$,
 which we do by Lemma~\ref{lem:general:affine:test}.
 
 Now assume there are at least two nonzero entries of $\hat{f}$ with distinct Hamming weights, say $\mathbf{u}_1, \mathbf{u}_2 \in \{0,1\}^n$.
 Then $\hat{f}_{\mathbf{u}_1}$ and $\hat{f}_{\mathbf{u}_2}$ are nonzero,
 and $0 < wt(\mathbf{u}_2) - wt(\mathbf{u}_1) \leq n$.
 Suppose there exists an $H = \tbmatrix{a}{b}{-b}{a} \in \mathbf{SO}_2(\mathbb{C})$ such that $g = H^{\otimes n} f \in \mathscr{A}$.
 Then by Lemma~\ref{lem:ortho:diag}, we have $\hat{g} = T^{\otimes n} \hat{f}$,
 where $T = \tbmatrix{a-bi}{0}{0}{a+bi}$ is a diagonal transformation.
 Recall $H_2$ and $D$ from Table \ref{tab:constants}.
 Since $Z' = \sqrt{2} H_2 D \in \StabA$, we have $\hat{g} = Z'^{\otimes n} g \in \mathscr{A}$.
 Also since $T$ is diagonal, both $\hat{g}_{\bf u_1}$ and $\hat{g}_{\bf u_2}$ are nonzero.
 Therefore, there must exist an $r \in \{0,1,2,3\}$ such that
 \begin{equation} \label{eqn:affine:single:entry_ratio}
  i^r
  = \frac{\hat{g}_{\bf u_2}} {\hat{g}_{\bf u_1}}
  = \frac{(a + b i)^{2 wt({\bf u_2}) - n} \hat{f}_{\bf u_2}} {(a + b i)^{2 wt({\bf u_1}) - n} \hat{f}_{\bf u_1}}
  = (a + b i)^{2 wt(\mathbf{u}_2) - 2 wt(\mathbf{u}_1)} \frac{\hat{f}_{\bf u_2}} {\hat{f}_{\bf u_1}},
 \end{equation}
 where we used \eqref{eqn:eiganvalue}.
 Recall that $0 < wt(\mathbf{u}_2) - wt(\mathbf{u}_1) \leq n$.
 View $a+bi$ as a variable, and then there are at most $2n$ solutions to \eqref{eqn:affine:single:entry_ratio}, given $r$ and $\hat{f}_{\bf u_1}$ and $\hat{f}_{\bf u_2}$.
 There are $4$ possible values of $r$,
 resulting in at most $8n$ many solutions for $a, b \in \mathbb{C}$ such that $a + b i$ satisfies~\eqref{eqn:affine:single:entry_ratio} and $a^2 + b^ 2 = 1$.
 Each $(a,b)$ solution corresponds to a distinct $H \in \mathbf{SO}_2(\mathbb{C})$.
\end{proof}

We also want to efficiently decide if there exists an $H \in \mathbf{SO}_2(\mathbb{C})$ such that $H^{\otimes n} f \in \tbmatrix{1}{0}{0}{\alpha} \mathscr{A}$.

\begin{lemma} \label{lem:general:affine:test:ortho_alpha}
 There is an algorithm to decide, for any input signature $f$ of arity $n$,
 whether there exists an $H \in \mathbf{SO}_2(\mathbb{C})$ such that $H^{\otimes n} f \in \tbmatrix{1}{0}{0}{\alpha} \mathscr{A}$ with running time polynomial in $N$.
 If so, either $f \in \tbmatrix{1}{0}{0}{\alpha} \mathscr{A}$ and $f$ is invariant under any transformation in $\mathbf{SO}_2(\mathbb{C})$,
 or there exist $O(n N^{16})$ many $H \in \mathbf{SO}_2(\mathbb{C})$ such that $H^{\otimes n} f \in \tbmatrix{1}{0}{0}{\alpha} \mathscr{A}$,
 and they can all be computed in polynomial time in $N$.
\end{lemma}

\begin{proof}
 Compute $\hat{f} = Z'^{\otimes n} f$.
 If the support of $\hat{f}$ contains at most one Hamming weight,
 then by Lemma~\ref{lem:ortho:invariant},
 $f$ is invariant under any $H \in \mathbf{SO}_2(\mathbb{C})$.
 Therefore we only need to directly decide if $f \in \tbmatrix{1}{0}{0}{\alpha} \mathscr{A}$,
 which we do by Corollary~\ref{cor:general:affine:test:alpha}.

 Now assume there are at least two nonzero entries of $\hat{f}$ that are of distinct Hamming weight.
 Let $\mathbf{u}_1, \mathbf{u}_2 \in \{0,1\}^n$ be such that $\hat{f}_{\mathbf{u}_1}$ and $\hat{f}_{\mathbf{u}_2}$ are nonzero,
 and $0 < wt(\mathbf{u}_2) - wt(\mathbf{u}_1) \leq n$.
 We derive necessary conditions for the existence of $H \in \mathbf{SO}_2(\mathbb{C})$ such that $H^{\otimes n} f \in \tbmatrix{1}{0}{0}{\alpha} \mathscr{A}$.
 Thus, assume such an $ H = \tbmatrix{a}{b}{-b}{a}$ exists, where $a^2 + b^2 = 1$.
 
 Let $g = H^{\otimes n} f$.
 Then $\hat{g} = Z'^{\otimes n} g \in \tbmatrix{1}{i}{1}{-i} \tbmatrix{1}{0}{0}{\alpha} \mathscr{A}$.
 By Lemma~\ref{lem:ortho:diag}, we have $\hat{g} = T^{\otimes n} \hat{f}$, where $T = \tbmatrix{a-bi}{0}{0}{a+bi}$.
 Thus $\hat{g}_\mathbf{u} = (a + b i)^{2 wt(\mathbf{u}) - n} \hat{f}_\mathbf{u}$ for any $\mathbf{u} \in \{0,1\}^n$.
 Let $t = wt(\mathbf{u}_1) - wt(\mathbf{u}_2)$.
 Then
 \[
  \frac{\hat{g}_{\mathbf{u}_1}}{\hat{g}_{\mathbf{u}_2}}
  = \frac{(a+bi)^{2wt(\mathbf{u}_1)-n}\hat{f}_{\mathbf{u}_1}} {(a+bi)^{2wt(\mathbf{u}_2)-n}\hat{f}_{\mathbf{u}_2}}
  = (a+bi)^{2t}\frac{\hat{f}_{\mathbf{u}_1}}{\hat{f}_{\mathbf{u}_2}}.
 \]
 Hence
 \[
  (a+bi)^{2t}
  = \frac{\hat{f}_{\mathbf{u}_2}}{\hat{f}_{\mathbf{u}_1}} \cdot \frac{\hat{g}_{\mathbf{u}_1}}{\hat{g}_{\mathbf{u}_2}}.
 \]

 We claim that the value of each entry in $\hat{g}$ as well as the number of possible values is bounded by a polynomial in $N$,
 and hence so are the ratios between them.
 Let $h \in \mathscr{A}$ be a signature such that $\hat{g} = \tbmatrix{1}{i}{1}{-i}^{\otimes n} \tbmatrix{1}{0}{0}{\alpha}^{\otimes n} h$.
 Every nonzero entry of $h$ is a power of $i$, up to a constant factor $\lambda$.
 This constant factor cancels when taking ratios of entries, so we omit it.
 Let $h' = \tbmatrix{1}{0}{0}{\alpha}^{\otimes n} h$.
 Then every entry of $h'$ is a power of $\alpha$ or~$0$.
 Moreover, each entry of $\tbmatrix{1}{i}{1}{-i}^{\otimes n}$ is also a power of $\alpha$.
 Therefore every entry of $\hat{g}$ is an exponential sum of $2^n$ terms, each a power of $\alpha$ or~$0$.
 Recall that $\alpha^8=1$ and hence there are $8$ possible values of these powers.
 Let $c_0$ denote the number of~$0$ and $c_i$ (for $1 \le i \le 8$) denote the number of $\alpha^i$ in an entry $\hat{g}_{\bf u}$ of $\hat{g}$.
 Then we have
 \[
  c_0 + \sum_{i=1}^8 c_i = 2^n
  \qquad
  \text{and}
  \qquad
  \sum_{i=1}^8 c_i \alpha^i = \hat{g}_{\bf u}.
 \]
 Clearly the total number of possible values of entries in $\hat{g}_{\bf u}$ is at most the number of possible choices of $(c_0, \ldots, c_8)$.
 There are at most $\binom{2^n + 8}{8} = O(N^8)$ choices of $(c_0, \ldots, c_8)$.
 Thus the number of all possible  ratios is at most $O(N^{16})$,
 and can all be enumerated in time polynomial in $N$.

 For any possible value of the ratio $\frac{\hat{g}_{\mathbf{u}_1}}{\hat{g}_{\mathbf{u}_2}}$,
 each possible value of $\frac{\hat{f}_{\mathbf{u}_2}}{\hat{f}_{\mathbf{u}_1}}$ gives at most $2n$ different transformations $H$.
 Therefore, the total number of transformations is bounded by $O(n N^{16})$,
 and we can find them in time polynomial in $N$.
\end{proof}

Now we give an algorithm that efficiently decides if a set of signatures is $\mathscr{A}$-transformable.

\begin{theorem} \label{thm:affine:decide}
 There is a polynomial-time algorithm to decide, for any finite set of signatures $\mathcal{F}$,
 whether $\mathcal{F}$ is $\mathscr{A}$-transformable.
 If so, at least one transformation can be found.
\end{theorem}

\begin{proof}
 By Lemma~\ref{lem:affine:trans:so2},
 we only need to decide if there exists an $H \in \mathbf{SO}_2(\mathbb{C})$ such that
 $\mathcal{F} \subseteq H \mathscr{A}$ or $\mathcal{F} \subseteq H \tbmatrix{1}{0}{0}{\alpha} \mathscr{A}$.
 To every signature in $\mathcal{F}$,
 we apply Lemma~\ref{lem:general:affine:test:ortho} or Lemma~\ref{lem:general:affine:test:ortho_alpha} to check each case, respectively.
 If no $H$ exists for some signature,
 then $\mathcal{F}$ is not $\mathscr{A}$-transformable.
 Otherwise, every signature is $\mathscr{A}$-transformable for some $H \in \mathbf{SO}_2(\mathbb{C})$.
 If every signature in $\mathcal{F}$ is invariant under transformations in $\mathbf{SO}_2(\mathbb{C})$,
 then $\mathcal{F}$ is $\mathscr{A}$-transformable.
 Otherwise, we pick the first $f \in \mathcal{F}$ that is not invariant under transformations in $\mathbf{SO}_2(\mathbb{C})$.
 The number of possible transformations that work for $f$ is bounded by a polynomial in the size of the presentation of $f$.
 We simply try all such transformations on all other signatures in $\mathcal{F}$ that are not invariant under transformations in $\mathbf{SO}_2(\mathbb{C})$,
 respectively using Lemma~\ref{lem:general:affine:test} or Corollary~\ref{cor:general:affine:test:alpha} to check if the transformation works.
\end{proof}

%% file: 5.General.Product.tex
\section{General \texorpdfstring{$\mathscr{P}$}{P}-transformable Signatures} \label{sec:general:product}

In this section, we give the algorithm to check $\mathscr{P}$-transformable signatures.
Once again, our general strategy is to bound the number of possible transformations (with a few exceptions), and then enumerate all of them.
Indeed, the bound will be a constant in this section.
The distinct feature for $\mathscr{P}$-transformable signatures is that we have to decompose them first.

We begin with the counterpart to Lemma~\ref{lem:affine:trans:so2},
which strengthens Lemma~\ref{lem:product:trans} by restricting to either
orthogonal transformations within $\mathbf{SO}_2(\mathbb{C})$ or no orthogonal transformation at all.

\begin{lemma} \label{lem:product:trans:so2}
 Let $\mathcal{F}$ be a set of signatures.
 Then $\mathcal{F}$ is $\mathscr{P}$-transformable iff $\mathcal{F} \subseteq \tbmatrix{1}{1}{i}{-i} \mathscr{P}$ or
 there exists an $H \in \mathbf{SO}_2(\mathbb{C})$ such that $\mathcal{F} \subseteq H \mathscr{P}$.
\end{lemma}

\begin{proof}
 Sufficiency is obvious by Lemma~\ref{lem:product:trans}.

 Assume that $\mathcal{F}$ is $\mathscr{P}$-transformable.
 By Lemma~\ref{lem:product:trans}, there exists an $H \in \mathbf{O}_2(\mathbb{C})$
 such that $\mathcal{F} \subseteq H \mathscr{P}$ or $\mathcal{F} \subseteq H \tbmatrix{1}{1}{i}{-i} \mathscr{P}$.
 There are two cases to consider.
 \begin{enumerate}
  \item Suppose $\mathcal{F} \subseteq H \mathscr{P}$.
  If $H \in \mathbf{SO}_2(\mathbb{C})$,
  then we are done,
  so assume that $H \in \mathbf{O}_2(\mathbb{C}) \setminus \mathbf{SO}_2(\mathbb{C})$.
  We want to find an $H' \in \mathbf{SO}_2(\mathbb{C})$ such that $\mathcal{F} \subseteq H' \mathscr{P}$.
  Let $H' = H \tbmatrix{1}{0}{0}{-1} \in \mathbf{SO}_2(\mathbb{C})$.
  Then
  \begin{align*}
   \mathcal{F}
   &\subseteq H \tbmatrix{1}{0}{0}{-1} \mathscr{P}\\
   &= H' \mathscr{P}
  \end{align*}
  since $\tbmatrix{1}{0}{0}{-1} \in \StabP$.
  
  \item Suppose $\mathcal{F} \subseteq H \tbmatrix{1}{1}{i}{-i} \mathscr{P}$.
  If $H = \tbmatrix{a}{b}{-b}{a} \in \mathbf{SO}_2(\mathbb{C})$,
  then
  \begin{align*}
   \mathcal{F}
   &\subseteq H \tbmatrix{1}{1}{i}{-i} \mathscr{P}\\
   &\subseteq \tbmatrix{1}{1}{i}{-i} \tbmatrix{a+bi}{0}{0}{a-bi} \mathscr{P}\\
   &\subseteq \tbmatrix{1}{1}{i}{-i} \mathscr{P}
  \end{align*}
  since $H \tbmatrix{1}{1}{i}{-i} = \tbmatrix{1}{1}{i}{-i} \tbmatrix{a+bi}{0}{0}{a-bi}$ and $\tbmatrix{a+bi}{0}{0}{a-bi} \in \StabP$.
  Otherwise, $H = \tbmatrix{a}{b}{b}{-a} \in \mathbf{O}_2(\mathbb{C}) \setminus \mathbf{SO}_2(\mathbb{C})$ and
  \begin{align*}
   \mathcal{F}
   &\subseteq H \tbmatrix{1}{1}{i}{-i} \mathscr{P}\\
   &\subseteq \tbmatrix{1}{1}{i}{-i} \tbmatrix{0}{a-bi}{a+bi}{0} \mathscr{P}\\
   &\subseteq \tbmatrix{1}{1}{i}{-i} \mathscr{P}
  \end{align*}
  since $H \tbmatrix{1}{1}{i}{-i} = \tbmatrix{1}{1}{i}{-i} \tbmatrix{0}{a-bi}{a+bi}{0}$ and $\tbmatrix{0}{a-bi}{a+bi}{0} \in \StabP$.
  \qedhere
 \end{enumerate}
\end{proof}

The ``building blocks'' of $\mathscr{P}$ are signatures whose support is contained in two entries with complementary indices.
However, for technical convenience that will be explained shortly,
in the following definition we restrict to functions that are either unary, or have support of size exactly two.
Recall that two signatures are considered the same if one is a nonzero multiple of the other.

\begin{definition} \label{def:E}
 A $k$-ary function $f$ is a \emph{generalized equality} if it is a nonzero multiple of $[0,0], [1,0]$, $[0,1]$, or satisfies
 \[
  \exists \mathbf{x} \in \{0,1\}^k,
  \quad
  \forall \mathbf{y} \in \{0,1\}^k,
  \quad
  f_\mathbf{y} = 0 \iff \mathbf{y} \not\in \{\mathbf{x}, \overline{\mathbf{x}}\}.
 \]
 We use $\mathscr{E}$ to denote the set of all generalized equality functions.
\end{definition}

For any set $\mathcal{F}$,
we let $\langle \mathcal{F} \rangle$ denote the closure under function products without shared variables.
It is easy to show that $\mathscr{P} = \langle \mathscr{E} \rangle$ (cf.~\cite{CLX11a}).

If we view signatures as tensors,
then $\langle \cdot \rangle$ is the closure under tensor products.
That is, if $f(\mathbf{x}_1, \mathbf{x}_2) = f_1(\mathbf{x}_1) f_2(\mathbf{x}_2)$,
then $f = f_1 \otimes f_2$ with a correct ordering of indices.
In general, we call such $f$ \emph{reducible}, defined next.

\begin{definition}\label{def:reducible}
 We call a function $f$ of arity $n$ on variable set $\mathbf{x}$ \emph{reducible}
 if $f$ has a non-trivial decomposition,
 namely, there exist $f_1$ and $f_2$ of arities $n_1$ and $n_2$ on variable sets $\mathbf{x}_1$ and $\mathbf{x}_2$, respectively,
 such that $1 \leq n_1, n_2 \leq n-1$, $\mathbf{x}_1 \union \mathbf{x}_2 = \mathbf{x}$,
 $\mathbf{x}_1 \intersect \mathbf{x}_2 = \emptyset$, and $f(\mathbf{x}) = f_1(\mathbf{x}_1) f_2(\mathbf{x}_2)$.
 Otherwise we call $f$ \emph{irreducible}.
\end{definition}

Note that all unary functions, including $[0,0]$, are irreducible.
However, the identically zero function of arity greater than one is reducible.
Recall that we call a function degenerate if it is a tensor product of unary functions.
All degenerate functions of arity $\ge 2$ are reducible, but not vice versa --- a reducible function may be decomposable into only non-unary functions.
Due to the same reason, degenerate functions are trivially tractable, but reducible functions are not necessarily so.

Definition~\ref{def:E} is a slight modification of a similar definition for $\mathscr{E}$ that appeared in Section~2 of~\cite{CLX11a}.
For both definitions of $\mathscr{E}$, it follows that $\mathscr{P} = \langle \mathscr{E} \rangle$.
The motivation for our slight change in the definition is so that every signature in $\mathscr{E}$ is irreducible.

Irreducibility is preserved by transformations.

\begin{lemma}\label{lem:reducible:transformation}
  Let $f$ be an irreducible function of arity $n$, and $T$ be a $2$-by-$2$ non-singular matrix.
  Then $g=T^{\otimes n}f$ is also irreducible.
\end{lemma}
\begin{proof}
  Suppose $g$ is reducible.
  By Definition \ref{def:reducible}, there is a non-trivial decomposition $g=g_1\otimes g_2$.
  Hence $f = \left(T^{-1}\right)^{\otimes n} g$ also has a non-trivial decomposition.
\end{proof}

If a function $f$ is reducible,
then we can factor it into functions of smaller arity.
This procedure can be applied recursively and terminates when all components are irreducible.
Therefore any function has at least one irreducible factorization.
We show that such a factorization is unique for functions that are not identically zero.

\begin{lemma} \label{lem:factorization:unique}
 Let $f$ be a function of arity $n$ on variables $\mathbf{x}$ that is not identically zero.
 Assume there exist irreducible functions $f_i$ and $g_j$,
 and two partitions $\{\mathbf{x}_i\}$ and $\{\mathbf{y}_j\}$ of $\mathbf{x}$ for $1 \leq i \leq k$ and $ 1 \leq j \leq k'$,
 such that
 \[
  f(\mathbf{x})
  = \prod_{i=1}^k    f_i(\mathbf{x}_i)
  = \prod_{j=1}^{k'} g_j(\mathbf{y}_j).
 \]
 Then $k = k'$, the partitions are the same, and there exists a permutation $\pi$ on $\{1,2,\cdots,k\}$ such that
 $f_i=g_{\pi(j)}$ up to nonzero factors.
\end{lemma}

\begin{proof}
 Since $f$ is not identically zero, none of the $f_i$ or $g_j$ is identically zero.
 Fix an assignment $u_2, \dotsc, u_k$ such that $c = \prod_{i=2}^k f_i(u_i) \neq 0$.
 Let $\mathbf{z}_j = \mathbf{y}_j \intersect \mathbf{x}_1$,
 and $\mathbf{v}_j = \mathbf{y}_j \intersect (\union_{i=2}^k \mathbf{x}_i)$ for $1 \leq j \leq k'$.
 Let the assignments $u_2, \dotsc, u_k$ restricted to $\mathbf{v}_j$ be $w_j$.
 Then we have
 \[
  c f_1(\mathbf{x}_1) 
  = f_1(\mathbf{x}_1) \prod_{i=2}^k f_i(u_i)
  = \prod_{j=1}^{k'} g_j(\mathbf{z}_j, w_j).
 \]
 Define new functions $h_j(\mathbf{z}_j) = g_j(\mathbf{z}_j, w_j)$ for $1 \leq j \leq k'$.
 Then 
 \[
  f_1(\mathbf{x}_1)
  = \frac{1}{c} \prod_{j=1}^{k'} h_j(\mathbf{z}_j).
 \]
 Since $f_1$ is irreducible, there cannot be two $\mathbf{z}_j$ that are nonempty.
 And yet, $\mathbf{x}_1 = \union_{j=1}^{k'} \mathbf{z}_j$,
 so it follows that $\mathbf{x}_1 = \mathbf{z}_j$ for some $1 \le j \le k'$.
 We may assume $j=1$, so $\mathbf{x}_1 \subseteq \mathbf{y}_1$.
 By the same argument we have $\mathbf{y}_1 \subseteq \mathbf{x}_i$, for some $i$.
 But by disjointness of $\mathbf{x} = \union_{i=1}^k \mathbf{x}_i$, we must have $\mathbf{y}_1 \subseteq \mathbf{x}_1$.
 Thus after a permutation, we have $\mathbf{x}_1 = \mathbf{y}_1$.
 Therefore $f_1 = g_1$ up to a nonzero constant.

 By fixing some assignment to $\mathbf{x}_1 = \mathbf{y}_1$ such that $f_1$ and $g_1$ are not zero,
 we may cancel this factor, and the proof is completed by induction.
 Therefore we must have that $k=k'$ and the two sets $\{f_i\}$ and $\{g_j\}$ are equal,
 where we identify functions up to nonzero constants.
\end{proof}

In fact, we can efficiently find the unique factorization.

\begin{lemma} \label{lem:factorization:find}
 There is an algorithm to compute in time polynomial in $N$, for any input signature $f$ of arity $n$ that is not identically zero,
 the unique factorization of $f$ into irreducible factors.
 More specifically, the algorithm computes irreducible $f_1, \dotsc, f_k$ of arities $n_1, \dotsc, n_k \in \mathbb{Z}^+$ (for some $k \ge 1$)
 such that $\sum_{i=1}^k n_i = n$ and $f(\mathbf{x_1}, \dotsc, \mathbf{x_k}) = \prod_{i=1}^k f_i(\mathbf{x_i})$.
\end{lemma}

\begin{proof}
 We may partition the variables $\mathbf{x}$ into two sets $\mathbf{x}_1$ and $\mathbf{x}_2$ of length $n_1$ and $n_2$, respectively,
 such that $1 \leq n_1, n_2 \leq n-1$, $\mathbf{x}_1 \union \mathbf{x}_2 = \mathbf{x}$, and $\mathbf{x}_1 \intersect \mathbf{x}_2 = \emptyset$.
 Define a $2^{n_1}$-by-$2^{n_2}$ matrix $M$ such that $M_{u_1, u_2} = f(u_1, u_2)$
 for $u_1 \in \{0,1\}^{n_1}$ and $u_2\in\{0,1\}^{n_2}$.
 Then $M$ is of rank at most~$1$ iff there exist $f_1$ and $f_2$ of arity $n_1$ and $n_2$,
 such that $f(\mathbf{x}) = f_1(\mathbf{x}_1) f_2(\mathbf{x}_2)$.

 Therefore, in order to factor $f$,
 we only need to run through all distinct partitions,
 and check if there exists at least one such matrix of rank at most~$1$.
 If none exists, then $f$ is irreducible.
 The total number of possible such partitions is $2^{n-1}-1$.
 Hence the running time is polynomial in $2^n \le N$.

 Once we have found $f = f_1 \otimes f_2$,
 we recursively apply the above procedure to $f_1$ and $f_2$ until every component is irreducible.
 The total running time is polynomial in $N$.
\end{proof}

This factorization algorithm gives a simple algorithm to determine membership in $\mathscr{P}$.

\begin{lemma} \label{lem:general:product:test}
 There is an algorithm to decide, for a given signature $f$ of arity $n$,
 whether $f \in \mathscr{P}$ with running time polynomial in $N$.
\end{lemma}

\begin{proof}
 We may assume that $f$ is not identically zero, and we obtain its unique factorization $f = \bigotimes_i f_i$ by Lemma~\ref{lem:factorization:find}.
 Then $f \in\mathscr{P}$ iff for all $i$, we have $f_i \in \mathscr{E}$.
 Since membership in $\mathscr{E}$ is easy to check, our proof is complete.
\end{proof}

Let $T \in \mathbf{GL}_2(\mathbb{C})$ be some transformation and $f$ some signature.
To check if $f \in T \mathscr{P}$,
it suffices to first factor $f$ and then check if each irreducible factor is in $T \mathscr{E}$.
 
\begin{lemma} \label{lem:product:test:trans}
 Suppose $f = \bigotimes_{i=1}^k f_i$ is not identically zero and that $f_i$ is irreducible for all $1 \leq i \leq k$.
 Let $T \in \mathbf{GL}_2(\mathbb{C})$.
 Then $f \in T \mathscr{P}$ iff $f_i \in T \mathscr{E}$ for all $1 \leq i \leq k$.
\end{lemma}

\begin{proof}
 Suppose $f$ is of arity $n$ and $f_i$ is of arity $n_i$ so that $\sum_{i=1}^k n_i = n$.
 If $f_i \in T \mathscr{E}$ for all $1 \leq i \leq k$,
 then there exists $g_i \in \mathscr{E}$ such that $f_i=T^{\otimes n_i} g_i$.
 Thus $f = \bigotimes_{i=1}^k f_i = \bigotimes_{i=1}^k T^{\otimes n_i} g_i = T^{\otimes n} \bigotimes_{i=1}^k g_i$.
 Since $g_i \in \mathscr{E}$, we have $\bigotimes_{i=1}^k g_i \in \mathscr{P}$.
 Therefore $f \in T\mathscr{P}$.

 On the other hand, assume $f \in T \mathscr{P}$.
 By the definition of $\mathscr{P}$,
 there exist $g_1, \dotsc, g_{k'} \in \mathscr{E}$ of arities $m_1, \dotsc, m_{k'} \in \mathbb{Z}^+$,
 such that $f = T^{\otimes n} g$, where $g = \bigotimes_{i=1}^{k'} g_i$.
 Since $g_i \in \mathscr{E}$, $g_i$ is irreducible.
 Let $f_i' = T^{\otimes m_i} g_i \in T \mathscr{E}$ for all $1 \leq i \leq k'$,
 which is also irreducible by Lemma \ref{lem:reducible:transformation}.
 Then $\bigotimes_{i=1}^{k'} f_i' = f = \bigotimes_{i=1}^k f_i$.
 By Lemma~\ref{lem:factorization:unique},
 we have $k = k'$ and $\{f_i\}$ and $\{f_i'\}$ are the same up to a permutation.
 Therefore each $f_i \in T \mathscr{E}$.
\end{proof}

With Lemma~\ref{lem:factorization:find} and Lemma~\ref{lem:product:test:trans} in mind,
we focus our attention on membership in $\mathscr{E}$.
We show how to efficiently decide if there exists an $H \in \mathbf{SO}_2(\mathbb{C})$ such that $H^{\otimes n} f \in \mathscr{E}$ when $f$ is irreducible.

\begin{lemma} \label{lem:product:ortho:test}
 There is an algorithm to decide, for a given irreducible signature $f$ of arity $n \geq 2$,
 whether there exists an $H \in \mathbf{SO}_2(\mathbb{C})$ such that $H^{\otimes n} f\in\mathscr{E}$ with running time polynomial in $N$.
 If so, there exist at most eight $H \in \mathbf{SO}_2(\mathbb{C})$
 such that $H^{\otimes n} f \in \mathscr{E}$ unless $f = \transpose{(1,0,0,1)}$ or $f = \transpose{(0,1,-1,0)}$.
\end{lemma}

\begin{proof}
 Assume there exists an $H = \tbmatrix{a}{b}{-b}{a} \in \mathbf{SO}_2(\mathbb{C})$ such that $g = H^{\otimes n} f \in \mathscr{E}$,
 where $a^2 + b^2 = 1$.
 Then by Lemma~\ref{lem:ortho:diag},
 there exists a diagonal transformation $T = \tbmatrix{a-bi}{0}{0}{a+bi}$ such that $\hat{g} = T^{\otimes n} \hat{f} \in \tbmatrix{1}{i}{1}{-i} \mathscr{E}$.
 In particular, $\hat{g}$ and $\hat{f}$ have the same support.
 For two vectors $\mathbf{u},\mathbf{x}\in\{0,1\}^{n}$,
 the entry indexed by row $\mathbf{u}$ and column $\mathbf{x}$ in the matrix $\tbmatrix{1}{i}{1}{-i}^{\otimes n}$
 is $i^{wt(\mathbf{x})} (-1)^{\langle \mathbf{x}, \mathbf{u} \rangle}$,
 where $wt(\cdot)$ denotes Hamming weight and $\langle \cdot, \cdot \rangle$ is the dot product.

 Since $g\in\mathscr{E}$, $g$ is irreducible.
 Thus $g$ has two nonzero entries with opposite index, say $\mathbf{x}$ and $\overline{\mathbf{x}}$.
 Hence we have
 \begin{align*}
  \hat{g}_{\mathbf{u}}
   &= i^{wt(\mathbf{x})}(-1)^{\langle\mathbf{x},\mathbf{u}\rangle} g_{\mathbf{x}}
    + i^{wt(\overline{\mathbf{x}})}(-1)^{\langle\overline{\mathbf{x}},\mathbf{u}\rangle} g_{\overline{\mathbf{x}}}\\
   &= i^{wt(\mathbf{x})}(-1)^{\langle\mathbf{x},\mathbf{u}\rangle} g_{\mathbf{x}}
    + i^{n-wt(\mathbf{x})}(-1)^{wt(\mathbf{u})-\langle\mathbf{x},\mathbf{u}\rangle} g_{\overline{\mathbf{x}}}\\
   &= (-1)^{\langle\mathbf{x},\mathbf{u}\rangle} \left(i^{wt(\mathbf{x})} g_{\mathbf{x}} + i^{n-wt(\mathbf{x})} (-1)^{wt(\mathbf{u})} g_{\overline{\mathbf{x}}}\right)
 \end{align*}
 for any vector $\mathbf{u} \in \{0,1\}^n$.

 For $\mathbf{u}_1, \mathbf{u}_2 \in \{0,1\}^n$,
 if $wt(\mathbf{u}_1) \equiv wt(\mathbf{u}_2) \pmod{2}$,
 then
 \begin{equation} \label{eqn:plus-minus-g-hat}
  \hat{g}_{\mathbf{u}_1} = \pm \hat{g}_{\mathbf{u}_2}.
 \end{equation}
 Therefore, if any entry of $\hat{f}$ with even Hamming weight is~$0$,
 then all entries with even Hamming weight are~$0$.
 This also holds for entries with odd Hamming weight.
 However, $\hat{f}$ is not identically zero because it is irreducible and of arity $n \ge 2$.
 Therefore, we know that either all entries of even Hamming weight are not~$0$ or all entries of odd Hamming weight are not~$0$.
 If $n \geq 3$, or if $n = 2$ and all entries of even Hamming weight are not~$0$,
 then we can take two nonzero entries of $\hat{f}$ whose Hamming weight differ by~$2$.
 Their ratio restricts the possible choices of $a+bi$,
 as in the proof of Lemma~\ref{lem:general:affine:test:ortho_alpha},
 because the only possible ratios for $\hat{g}_{\mathbf{u}_1} / \hat{g}_{\mathbf{u}_2}$ are $\pm 1$ by~(\ref{eqn:plus-minus-g-hat}).
 Together with $a^2 + b^2 = 1$,
 this gives at most~$8$ possible matrices $H \in \mathbf{SO}_2(\mathbb{C})$.

 The remaining case is when $n=2$ and all entries of $\hat{f}$ with even Hamming weight are~$0$.
 By~(\ref{eqn:plus-minus-g-hat}),
 we have $\hat{g} = \lambda \transpose{(0,1,\pm 1,0)}$ for some $\lambda \not = 0$ since $\hat{g}$ and $\hat{f}$ have the same support.
 Then from $\hat{f} = (T^{-1})^{\otimes 2} \hat{g}$,
 where $T^{-1} = \tbmatrix{a+bi}{0}{0}{a-bi}$ is diagonal,
 we calculate that $T^{-1} \tbmatrix{0}{1}{\pm 1}{0} \transpose{(T^{-1})} = \tbmatrix{0}{1}{\pm 1}{0}$.
 Hence, up to a nonzero scalar, $\hat{f} = \transpose{(0,1,1,0)}$ or $\hat{f} = \transpose{(0,1,-1,0)}$.
 Finally $f = (Z'^{-1})^{\otimes 2} \hat{f}$, and we get
 $f = \transpose{(1,0,0,1)}$ or $f = \transpose{(0,1,-1,0)}$, up to a nonzero scalar.
\end{proof}

Now we give an algorithm that efficiently decides if a set of signatures is $\mathscr{P}$-transformable.

\begin{theorem} \label{thm:product:decide}
 There is a polynomial-time algorithm to decide,
 for any finite set of signatures $\mathcal{F}$,
 whether $\mathcal{F}$ is $\mathscr{P}$-transformable.
 If so, at least one transformation can be found.
\end{theorem}

\begin{proof}
 By Lemma~\ref{lem:product:trans:so2},
 we only need to decide if $\mathcal{F} \subseteq \tbmatrix{1}{1}{i}{-i} \mathscr{P}$ or
 if there exists an $H \in \mathbf{SO}_2(\mathbb{C})$ such that $\mathcal{F} \subseteq H \mathscr{P}$.
 To check if $\mathcal{F} \subseteq \tbmatrix{1}{1}{i}{-i} \mathscr{P}$,
 we simply apply Lemma~\ref{lem:general:product:test} to each signature in $\tbmatrix{1}{1}{i}{-i}^{-1} \mathcal{F}$.
 
 Now to check if $\mathcal{F} \subseteq H \mathscr{P}$.
 We may assume that no signature in $\mathcal{F}$ is identically zero.
 Now we obtain the unique factorization of each signature in $\mathcal{F}$ using Lemma~\ref{lem:factorization:find}.
 If every irreducible factor is either a unary signature, or $\transpose{(1,0,0,1)}$, or $\transpose{(0,1,-1,0)}$,
 then $\mathcal{F} \subseteq \langle \mathscr{E} \rangle = \mathscr{P}$.
 Otherwise, let $f \in \mathcal{F}$ be a signature that is not of this form.
 This means that $f$ has a unique factorization $f = \bigotimes_i f_i$ 
 where some $f_i$ is not a unary signature, or $\transpose{(1,0,0,1)}$, or $\transpose{(0,1,-1,0)}$.
 Assume it is $f_1$.
  
 By applying Lemma~\ref{lem:product:test:trans} to $f$,
 we get the necessary condition $f_1 \in H \mathscr{E}$.
 Then we apply Lemma~\ref{lem:product:ortho:test} to $f_1$.
 If the test passes, then by the definition of $f_1$,
 we have at most eight transformations in $\mathbf{SO}_2(\mathbb{C})$ that could work.
 For each possible transformation $H$,
 we apply Lemma~\ref{lem:general:product:test} to every signature in $H^{-1} \mathcal{F}$ to check if it works.
\end{proof}

%% file: 6.1.Symmetric.Affine.Single.tex
\section{Symmetric \texorpdfstring{$\mathscr{A}$}{A}-transformable Signatures} \label{sec:symmetric:affine}

In the next two sections, we consider the case when the signatures are symmetric.
The significant difference is that a symmetric signature of arity $n$ is given by $n + 1$ values, instead of $2^n$ values.
This exponentially more succinct representation requires us to find a more efficient algorithm.

\subsection{A Single Signature}

Recall Definition~\ref{def:A123}.
To begin with, we provide efficient algorithms to decide membership in each of $\mathscr{A}_1$, $\mathscr{A}_2$, and $\mathscr{A}_3$ for a single signature.
If the signature is in one of the sets,
then the algorithm also finds at least one corresponding orthogonal transformation satisfying Definition~\ref{def:A123}.
By Lemma~\ref{lem:cha:affine}, this is enough to check if a single signature is $\mathscr{A}$-transformable.

We say a signature $f$ satisfies a second order recurrence relation, 
if there exist not all zero $a, b, c \in \mathbb{C}$,
such that for all $0 \leq k \leq n-2$, $a f_k + b f_{k+1} + c f_{k+2} = 0$.
For a non-degenerate signature of arity at least~$3$,
these coefficients are unique up to a nonzero scalar.

\begin{lemma} \label{lem:symmetric:2nd_ord_rec_rel:unique}
 Let $f$ be a non-degenerate symmetric signature of arity $n \ge 3$.
 If $f$ satisfies a second order recurrence relation with coefficients $a,b,c \in \mathbb{C}$ and another one with coefficients $a',b',c' \in \mathbb{C}$,
 then there exists a nonzero $k \in \mathbb{C}$ such that $(a,b,c) = k (a',b',c')$.
\end{lemma}

\begin{proof}
 A function $f = [f_0, f_1, \dotsc, f_n]$ is degenerate if and only if $f_0,\dotsc,f_n$ forms a geometric sequence.
 As $f$ is non-degenerate,
 the matrix $A = \left[\begin{smallmatrix} f_0 & f_1 & \ldots & f_{n-1} \\ f_1 & f_2 & \ldots & f_n \end{smallmatrix}\right]$ has rank~$2$.
 Let $B = \left[\begin{smallmatrix} f_0 & f_1 & \ldots & f_{n-2} \\ f_1 & f_2 & \ldots & f_{n-1} \\ f_2 & f_3 & \ldots & f_{n} \end{smallmatrix}\right]$.
 We claim that $\rank(B) \ge 2$, which implies that $f$ satisfies at most one second order recurrence relation up to a nonzero scalar, as desired.
 
 If $(f_1, \dotsc, f_{n-1}) = \mathbf{0}$, then $f_0, f_n \neq 0$ since $\rank(A) = 2$, so $\rank(B) = 2$ as well.
 Otherwise, $(f_1, \ldots, f_{n-1}) \neq \mathbf{0}$.
 Consider the matrices
 $A_1 = \left[\begin{smallmatrix} f_0 & f_1 & \ldots & f_{n-2} \\ f_1 & f_2 & \ldots & f_{n-1} \end{smallmatrix}\right]$ and
 $A_2 = \left[\begin{smallmatrix} f_1 & f_2 & \ldots & f_{n-1} \\ f_2 & f_3 & \ldots & f_{n} \end{smallmatrix}\right]$,
 which are submatrices of both $A$ and $B$.
 Both $A_1$ and $A_2$ have rank at least~$1$ since $(f_1, \ldots, f_{n-1}) \neq \mathbf{0}$.
 We show that either $\rank(A_1) = 2$ or $\rank(A_2) = 2$, which implies that $\rank(B) \ge 2$.

 For a contradiction, suppose $\rank(A_1) = \rank(A_2) = 1$.
 Then there exist $\lambda, \mu \in \mathbb{C}$ such that
 $(f_0, \ldots, f_{n-2}) = \lambda (f_1, \ldots, f_{n-1})$ and $(f_2, \ldots, f_{n}) = \mu (f_1, \ldots, f_{n-1})$.
 If $\lambda = 0$, then $f_0=f_1=0$ as $n\ge 3$.
 It implies that $\rank(A_2) = \rank (A)$.
 However, $\rank(A) = 2$, a contradiction.
 Similarly if $\mu = 0$, then $\rank(A_1) = 2$, a contradiction.
 Otherwise $\lambda, \mu \neq 0$ and we get $f_i \neq 0$ for all $0 \le i \le n$, and $\lambda \mu = 1$.
 This implies that $\rank(A) = 1$, a contradiction.
\end{proof}




For a signature with a second order recurrence relation,
the quantity $b^2 - 4 a c$ is nonzero precisely when the signature can be expressed as the sum of two degenerate signatures that are linearly independent.

\begin{lemma} \label{lem:symmetric:2nd_ord_rec_rel:characterization}
 Let $f$ be a non-degenerate symmetric signature of arity $n \ge 3$.
 Then $f$ satisfies a second order recurrence relation with coefficients $a,b,c$ satisfying $b^2 - 4 a c \ne 0$
 iff there exist $a_0, b_0, a_1, b_1$ (satisfying $a_0 b_1 \ne a_1 b_0$) such that $f = \tbcolvec{a_0}{b_0}^{\otimes n} + \tbcolvec{a_1}{b_1}^{\otimes n}$.
\end{lemma}

\begin{proof}
  The ``only if'' direction is straightforward to verify.
  For the other direction, assume that there exist $a, b, c \in \mathbb{C}$ not all zero,
  such that for all $0 \leq k \leq n-2$, $a f_k + b f_{k+1} + c f_{k+2} = 0$.
  If $c\neq 0$, then since $b^2-4ac\neq 0$, we can solve this recurrence with the initial values of $f_0$ and $f_1$,
  namely, there exist $c_0,c_1\neq 0$ and $\lambda_1\neq\lambda_2$ such that for any $0\le k\le n$,
  \begin{align*}
    f_k = c_0 \lambda_1^k + c_1\lambda_2^k.
  \end{align*}
  In other words, we can express $f$ as $f = c_0\tbcolvec{1}{\lambda_1}^{\otimes n} + c_1\tbcolvec{1}{\lambda_2}^{\otimes n}$.
  Normalizing shows the claim.

  The other case of $c=0$ implies that $b\neq 0$. 
  Hence the entries $f_0,\cdots,f_{n-1}$ satisfy a first order recurrence relation and the recurrence does not involve the last entry $f_n$.
  Thus there must exist $c_0,c_1$ and $\lambda$ such that $f=c_0\tbcolvec{1}{\lambda}^{\otimes n} + c_1\tbcolvec{0}{1}^{\otimes n}$.
  Moreover, if any of $c_0$ or $c_1$ equals $0$, then $f$ is degenerate which contradicts the assumption.
  The lemma follows from a normalization.
\end{proof}

The following definition of the $\theta$ function is crucial.
A priori, $\theta(v_0, v_1)$ may be not well-defined, 
but this is circumvented by insisting that $v_0$ and $v_1$ be linearly independent.

\begin{definition} \label{def:sym:theta}
 For a pair of linearly independent vectors $v_0 = \tbcolvec{a_0}{b_0}$ and $v_1 = \tbcolvec{a_1}{b_1}$,
 we define
 \[
  \theta(v_0, v_1) := \left(\frac{a_0 a_1 + b_0 b_1}{a_1 b_0 - a_0 b_1}\right)^2.
 \]
 Furthermore, suppose that a signature $f$ of arity $n \ge 3$ can be expressed as $f = v_0^{\otimes n} + v_1^{\otimes n}$,
 where $v_0$ and $v_1$ are linearly independent.
 Then we define $\theta(f) = \theta(v_0, v_1)$.
\end{definition}

Intuitively, this formula is the square of the cotangent of the angle from $v_0$ to $v_1$.
This notion of cotangent is properly extended to the complex domain.
The expression is squared so that $\theta(v_0, v_1) = \theta(v_1, v_0)$.

Let $f = v_0^{\otimes n} + v_1^{\otimes n}$ be a non-degenerate signature of arity $n \ge 3$.
Since $f$ is non-degenerate, $v_0$ and $v_1$ are linearly independent.
The next proposition implies that this expression for $f$ via $v_0$ and $v_1$ is unique up to a root of unity.
Therefore, $\theta(f)$ from Definition~\ref{def:sym:theta} is well-defined.

\begin{proposition}[Lemma~9.1 in~\cite{CLX11c}] \label{prop:vector:uniq}
 Let $\mathbf{a}, \mathbf{b}, \mathbf{c}, \mathbf{d}$ be four vectors and suppose that $\mathbf{c}, \mathbf{d}$ are linearly independent.
 If for some $n \ge 3$, we have $\mathbf{a}^{\otimes n} + \mathbf{b}^{\otimes n} = \mathbf{c}^{\otimes n} + \mathbf{d}^{\otimes n}$,
 then there exist $\omega_0$ and $\omega_1$ satisfying $\omega_0^n = \omega_1^n = 1$ such that either
 $\mathbf{a} = \omega_0 \mathbf{c}$ and $\mathbf{b} = \omega_1 \mathbf{d}$ or
 $\mathbf{a} = \omega_0 \mathbf{d}$ and $\mathbf{b} = \omega_1 \mathbf{c}$.
\end{proposition}

For the convenience of future use, we can generalize Proposition~\ref{prop:vector:uniq} to the following simple lemma.

\begin{lemma} \label{lem:vector:uniq}
 Let $\mathbf{a}, \mathbf{b}, \mathbf{c}, \mathbf{d}$ be four vectors and suppose that $\mathbf{c}, \mathbf{d}$ are linearly independent.
 Furthermore, let $x_0,x_1,y_0,y_1$ be nonzero scalars.
 If for some $n \ge 3$, we have $x_0 \mathbf{a}^{\otimes n} + x_1 \mathbf{b}^{\otimes n} = y_0 \mathbf{c}^{\otimes n} + y_1 \mathbf{d}^{\otimes n}$,
 then there exist $\omega_0$ and $\omega_1$,  such that either
 $\mathbf{a} = \omega_0 \mathbf{c}$, $\mathbf{b} = \omega_1 \mathbf{d}$, $x_0 \omega_0^n = y_0$, and $x_1 \omega_1^n = y_1$; or
 $\mathbf{a} = \omega_0 \mathbf{d}$, $\mathbf{b} = \omega_1 \mathbf{c}$, $x_0 \omega_0^n = y_1$, and $x_1 \omega_1^n = y_0$.
\end{lemma}

It is easy to verify that $\theta$ is invariant under an orthogonal transformation.

\begin{lemma} \label{lem:symmetric:angle:ortho_invariant}
 For two linearly independent vectors $v_0$, $v_1 \in \mathbb{C}^2$ and $H \in \mathbf{O}_2(\mathbb{C})$,
 let $\widehat{v_0} = H v_0$ and $\widehat{v_1} = H v_1$.
 Then $\theta(v_0, v_1) = \theta(\widehat{v_0}, \widehat{v_1})$.
\end{lemma}

\begin{proof}
 Within the square in the definition of $\theta$,
 the numerator is the dot product, which is invariant under any orthogonal transformation.
 Also, the denominator is the determinant,
 which is invariant under any orthogonal transformation up to a sign.
\end{proof}

Now we have some necessary conditions for membership in $\mathscr{A}_1 \cup \mathscr{A}_2 \cup \mathscr{A}_3$.
Recall that $\mathscr{A}_1 \subseteq \mathscr{P}_1$.

\begin{lemma} \label{lem:symmetric:angle:AP-trans}
 Let $f$ be a non-degenerate symmetric signature of arity at least~$3$.
 Then
 \begin{enumerate}
  \item $f \in \mathscr{P}_1 \implies \theta(f) = 0$,
  \item $f \in \mathscr{A}_2 \implies \theta(f) = -1$, and
  \item $f \in \mathscr{A}_3 \implies \theta(f) = -\frac{1}{2}$.
 \end{enumerate}
\end{lemma}

\begin{proof}
 The result clearly holds when $f$ is in the canonical form of each set.
 This extends to the rest of each set by Lemma~\ref{lem:symmetric:angle:ortho_invariant}.
\end{proof}

These results imply the following corollary.

\begin{corollary}
 Let $f$ be a non-degenerate symmetric signature $f$ of arity $n \geq 3$.
 If $f$ is $\mathscr{A}$-transformable,
 then $f$ is of the form $v_0^{\otimes n} + v_1^{\otimes n}$,
 where $v_0$ and $v_1$ are linearly independent, and $\theta(v_0, v_1) \in \{0, -1, -\frac{1}{2}\}$.
\end{corollary}

The condition given in Lemma~\ref{lem:symmetric:angle:AP-trans} is not sufficient to determine if $f \in \mathscr{A}_1 \union \mathscr{A}_2 \union \mathscr{A}_3$.
For example, if $f = v_0^{\otimes n} + v_1^{\otimes n}$ with $v_0 = \tbcolvec{1}{i}$ and $v_1$ is not a multiple of $\tbcolvec{1}{-i}$,
then $\theta(f) = -1$ but $f$ is not in $\mathscr{A}_2$.
However, this is essentially the only exceptional case.
We achieve the full characterization with some extra conditions.

The next lemma gives an equivalent form for membership in $\mathscr{A}_1$, $\mathscr{A}_2$, and $\mathscr{A}_3$ using transformations in $\mathbf{O}_2(\mathbb{C}) \setminus \mathbf{SO}_2(\mathbb{C})$.
Only having to consider transformation matrices in $\mathbf{O}_2(\mathbb{C}) \setminus \mathbf{SO}_2(\mathbb{C})$ is convenient since such matrices are their own inverses.

\begin{lemma} \label{lem:o2-so2}
 Suppose $f$ is a non-degenerate symmetric signature of arity $n \geq 3$ and let $\mathscr{F} \in \{\mathscr{A}_1, \mathscr{A}_2, \mathscr{A}_3\}$.
 Then $f \in \mathscr{F}$ iff there exists an $H \in \mathbf{O}_2(\mathbb{C}) \setminus \mathbf{SO}_2(\mathbb{C})$ such that $f \in \mathscr{F}$ with $H$.
\end{lemma}

\begin{proof}
 Sufficiency is trivial.
 For necessity, assume that $f \in \mathscr{F}$ with $H \in \mathbf{O}_2(\mathbb{C})$.
 If $H \in \mathbf{O}_2(\mathbb{C}) \setminus \mathbf{SO}_2(\mathbb{C})$, then we are done,
 so further assume that $H \in \mathbf{SO}_2(\mathbb{C})$.
 By the definition of $\mathscr{F}$,
 \[f = c H^{\otimes n} \left(v_0^{\otimes n} + \beta v_1^{\otimes n}\right),\]
 where $c \neq 0$ and $v_0$, $v_1$, and $\beta$ depend on $\mathscr{F}$.
 Let $H' = \tbmatrix{1}{0}{0}{-1} H^{-1} \in \mathbf{O}_2(\mathbb{C}) \setminus \mathbf{SO}_2(\mathbb{C})$, so it follows that $\transpose{H{'}} = H'^{-1} = H'$.
 Then
 \begin{align*}
  f
  &= (H' H')^{\otimes n} f\\
  &= c H'^{\otimes n} (H' H)^{\otimes n} \left(v_0^{\otimes n} + \beta v_1^{\otimes n}\right)\\
  &= c H'^{\otimes n} \tbmatrix{1}{0}{0}{-1}^{\otimes n} \left(v_0^{\otimes n} + \beta v_1^{\otimes n}\right)\\
  &= c H'^{\otimes n} \left(v_1^{\otimes n} + \beta v_0^{\otimes n}\right)\\
  &= c \beta H'^{\otimes n} \left(v_0^{\otimes n} + \beta^{-1} v_1^{\otimes n}\right),
 \end{align*}
 where in the fourth step, we use the fact that $\tbmatrix{1}{0}{0}{-1} v_0 = v_1$ and $\tbmatrix{1}{0}{0}{-1} v_1 = v_0$
 for any $\mathscr{F} \in \{\mathscr{A}_1, \mathscr{A}_2, \mathscr{A}_3\}$.
 To finish, we rewrite $\beta^{-1}$ in the form required in Definition \ref{def:A123} as follows:
 \begin{itemize}
  \item if $\mathscr{F} = \mathscr{A}_1$, then $\beta = \alpha^{t n + 2 r}$ for some $t \in \{0,1\}$ and $r \in \{0,1,2,3\}$ and $\beta^{-1} = \alpha^{-t n - 2 r}$.
  Pick $r' \in \{0,1,2,3\}$ such that $r' \equiv -t n - r \pmod{4}$, so $\beta^{-1} = \alpha^{t n + 2 r'}$ as required;
  \item if $\mathscr{F} = \mathscr{A}_2$, then $\beta = 1$, so $\beta^{-1} = 1 = \beta$ as required;
  \item if $\mathscr{F} = \mathscr{A}_3$, then $\beta = i^{r}$ for some $r \in \{0,1,2,3\}$, so $\beta^{-1} = i^{-r} = i^{4 - r}$ as required. \qedhere
 \end{itemize}
\end{proof}

Before considering $\mathscr{A}_1$,
we prove a technical lemma that is also applicable when considering $\mathscr{P}_1$.

\begin{lemma} \label{lem:single:theta=0:H-less_form}
 Let $f = v_0^{\otimes n} + v_1^{\otimes n}$ be a symmetric signature of arity $n \geq 3$,
 where $v_0 = \tbcolvec{a_0}{b_0}$ and $v_1 = \tbcolvec{a_1}{b_1}$ are linearly independent.
 If $\theta(f) = 0$,
 then there exist an $H \in \mathbf{O}_2(\mathbb{C})$ and a nonzero $k \in \mathbb{C}$ satisfying $a_1 = k b_0$ and $b_1 = -k a_0$ such that
 \[
  H^{\otimes n} f = \lambda \left(\tbcolvec{1}{ 1}^{\otimes n}
                             + k^n \tbcolvec{1}{-1}^{\otimes n}\right)
 \]
 for some nonzero $\lambda \in \mathbb{C}$.
\end{lemma}

\begin{proof}
 Since $\theta(f) = 0$,
 we have $a_0 a_1 + b_0 b_1 = 0$.
 By linear independence, we have $a_1 b_0 \neq a_0 b_1$.
 Thus, there exists a nonzero $k \in \mathbb{C}$ such that $a_1 = k b_0$ and $b_1 = -k a_0$.
 (Note that this is clearly true even if one of $a_0$ or $b_0$, but not both, is zero.)
 Let $c = a_0^2 + b_0^2$, which is nonzero since $a_1 b_0 \neq a_0 b_1$.
 Also, let $u_0 = \frac{v_0}{\sqrt{c}}$ and $u_1 = \frac{v_1}{k \sqrt{c}}$, so it follows that the matrix $M = [u_0\ u_1]$ is orthogonal.
 Then the matrix $H = \frac{1}{\sqrt{2}} \tbmatrix{1}{1}{1}{-1} M^{-1}$ is also orthogonal and what we need.
 Under a transformation by $H$, we have
 \begin{align*}
  H^{\otimes n}f
  &= H^{\otimes n} \left(c^{\frac{n}{2}} u_0^{\otimes n} + k^n c^{\frac{n}{2}} u_1^{\otimes n}\right)\\
  &= \lambda \left(\tbcolvec{1}{1}^{\otimes n} + k^n \tbcolvec{1}{-1}^{\otimes n}\right),
 \end{align*}
 where $\lambda = (c/2)^{\frac{n}{2}} \neq 0$.
\end{proof}

Now we give the characterization of $\mathscr{A}_1$.

\begin{lemma} \label{lem:single:A1}
 Let $f = v_0^{\otimes n} + v_1^{\otimes n}$ be a symmetric signature of arity $n \geq 3$,
 where $v_0 = \tbcolvec{a_0}{b_0}$ and $v_1 = \tbcolvec{a_1}{b_1}$ are linearly independent.
 Then $f \in \mathscr{A}_1$ iff $\theta(f) = 0$ and there exist an $r \in \{0,1,2,3\}$ and $t \in \{0,1\}$
 such that $a_1^n = \alpha^{t n + 2 r} b_0^n \neq 0$ or $b_1^n = \alpha^{t n + 2 r} a_0^n \neq 0$.
\end{lemma}

\begin{proof}
 Suppose $f \in \mathscr{A}_1$.
 By Lemma~\ref{lem:o2-so2}, after a suitable normalization,
 there exists a transformation $H = \tbmatrix{x}{y}{y}{-x} \in \mathbf{O}_2(\mathbb{C}) \setminus \mathbf{SO}_2(\mathbb{C})$ such that
 \[
  f = H^{\otimes n} \left(\tbcolvec{1}{ 1}^{\otimes n}
                  + \beta \tbcolvec{1}{-1}^{\otimes n}\right),
 \]
 where $\beta = \alpha^{t n + 2 r}$ for some $r \in \{0,1,2,3\}$ and some $t \in \{0,1\}$.
 Since $H \in \mathbf{O}_2(\mathbb{C})$, we have $x^2 + y^2 = 1$.
 By Lemma~\ref{lem:symmetric:angle:AP-trans}, $\theta(f) = 0$.

 Now we have two expressions for $f$, which are
 \[
  \tbcolvec{a_0}{b_0}^{\otimes n} + \tbcolvec{a_1}{b_1}^{\otimes n}
  = f
  = \tbcolvec{x+y}{y-x}^{\otimes n} + \beta \tbcolvec{x-y}{y+x}^{\otimes n}.
 \]
 Since $v_0$ and $v_1$ are linearly independent,
 we know that $a_0$ and $a_1$ cannot both be $0$.
 Suppose $a_0 \neq 0$.
 By Lemma~\ref{lem:vector:uniq}, we have two cases.
 \begin{enumerate}
  \item Suppose $a_0 = \omega_0 (x+y)$ and $b_1 = \omega_1 (x+y)$ where $\omega_0^n = 1$ and $\omega_1^n = \beta$.
  Then we have $b_1^n = \beta (x+y)^n = \beta a_0^n \neq 0$.
  Since $\beta = \alpha^{t n + 2 r}$, we are done.
  \item Suppose $a_0 = \omega_0 (x-y)$ and $b_1 = \omega_1 (y-x)$ where $\omega_0^n = \beta$ and $\omega_1^n = 1$.
  Then we have $a_0^n = \beta (x-y)^n = \alpha^{t n + 2 r} (-1)^n (y-x)^n = \alpha^{t n + 2 r + 4 n} b_1^n$,
  so $b_1^n = \alpha^{-t n - 2 r - 4 n} a_0^n \neq 0$.
  Pick $r' \in \{0,1,2,3\}$ such that $r' \equiv -t n - r - 2 n \pmod{4}$.
  Then $\alpha^{-t n - 2 r - 4 n} = \alpha^{t n + 2 r'}$ is of the desired form.
 \end{enumerate}
 Otherwise, $a_1 \ne 0$, in which case, similar reasoning shows that $a_1^n = \alpha^{t n + 2 r} b_0^n \neq 0$.
 
 For sufficiency, we apply Lemma~\ref{lem:single:theta=0:H-less_form}, which gives
 \[
  H^{\otimes n} f = \lambda \left(\tbcolvec{1}{1}^{\otimes n} + k^n \tbcolvec{1}{-1}^{\otimes n}\right)
 \]
 for some $H \in \mathbf{O}_2(\mathbb{C})$,
 some nonzero $\lambda \in \mathbb{C}$,
 and some nonzero $k \in \mathbb{C}$ satisfying $a_1 = k b_0$ and $b_1 = -k a_0$.
 The ratio of these coefficients is $k^n$.
 We consider two cases.
 \begin{enumerate}
  \item Suppose $a_1^n = \alpha^{t n + 2 r} b_0^n \neq 0$.
  Then $k^n = \alpha^{t n + 2 r}$, so $f \in \mathscr{A}_1$.
  \item Suppose $b_1^n = \alpha^{t n + 2 r} a_0^n \neq 0$.
  Then $k^n = (-1)^n \alpha^{t n + 2 r}$.
  Pick $r' \in \{0,1,2,3\}$ such that $r' \equiv r + 2 n \pmod{4}$.
  Then $k^n = \alpha^{t n + 2 r'}$, so $f \in \mathscr{A}_1$.
  \qedhere
 \end{enumerate}
\end{proof}

Now we give the characterization of $\mathscr{A}_3$.

\begin{lemma} \label{lem:single:A3}
 Let $f = v_0^{\otimes n} + v_1^{\otimes n}$ be a symmetric signature of arity $n \geq 3$,
 where $v_0 = \tbcolvec{a_0}{b_0}$ and $v_1 = \tbcolvec{a_1}{b_1}$ are linearly independent.
 Then $f \in \mathscr{A}_3$ iff there exist an $\varepsilon \in \{1,-1\}$ and $r \in \{0,1,2,3\}$ such that
 $a_1 \left(\sqrt{2} a_0 + \varepsilon i b_0\right) = b_1 \left(\varepsilon i a_0 - \sqrt{2} b_0\right)$,
 $a_1^n = i^r \left(\varepsilon i a_0 - \sqrt{2} b_0\right)^n$,
 and $b_1^n = i^r \left(\sqrt{2} a_0 + \varepsilon i b_0\right)^n$.
\end{lemma}

\begin{proof}
 Suppose $f \in \mathscr{A}_3$.
 By Lemma~\ref{lem:o2-so2}, after a suitable normalization,
 there exists a transformation $H = \tbmatrix{x}{y}{y}{-x} \in \mathbf{O}_2(\mathbb{C})-\mathbf{SO}_2(\mathbb{C})$ such that
 \[
  f = H^{\otimes n} \left(\tbcolvec{1}{ \alpha}^{\otimes n}
                    + i^r \tbcolvec{1}{-\alpha}^{\otimes n}\right)
 \]
 for some $r \in \{0,1,2,3\}$.
 Since $H \in \mathbf{O}_2(\mathbb{C})$, we have $x^2 + y^2 = 1$.
 By Lemma~\ref{lem:symmetric:angle:AP-trans}, $\theta(f) = -\frac{1}{2}$,
 which implies $\frac{a_0 a_1 + b_0 b_1}{a_0 b_1 - a_1 b_0} = \pm \frac{i}{\sqrt{2}}$.
 After rearranging terms, we get
 \[
  a_1 \left(\sqrt{2} a_0 + \varepsilon i b_0\right) = b_1 \left(\varepsilon i a_0 - \sqrt{2} b_0\right),
 \]
 for some $\varepsilon \in \{1,-1\}$.
 Since $v_0$ and $v_1$ are linearly independent,
 we know that $a_1$ and $b_1$ cannot both be $0$.
 Also, if $\sqrt{2} a_0 + \varepsilon i b_0$ and $\varepsilon i a_0 - \sqrt{2} b_0$ are both $0$,
 then we have $-\sqrt{2} a_0 = \varepsilon i b_0$ and $\varepsilon i a_0 = \sqrt{2} b_0$,
 which implies $a_0 = b_0 = 0$, a contradiction.
 Therefore, we have 
 \begin{equation} \label{eqn:a1b1-expression}
  a_1 = c (\varepsilon i a_0 - \sqrt{2} b_0)
  \qquad \text{and} \qquad
  b_1 = c (\sqrt{2} a_0 + \varepsilon i b_0)
 \end{equation}
 for some $c \neq 0$.
 To prove necessity, it remains to show that $c^n$ is a power of $i$.

 Now using $H^{-1} = H$, we have two expressions for $(H^{-1})^{\otimes n} f$, which are
 \[
  \tbcolvec{x a_0+y b_0}{y a_0 - x b_0}^{\otimes n} + \tbcolvec{ x a_1 + y b_1}{ y a_1 - x b_1}^{\otimes n}
  = H^{\otimes n} \left(\tbcolvec{a_0}{b_0}^{\otimes n} + \tbcolvec{a_1}{b_1}^{\otimes n}\right)
  = \left(H^{-1}\right)^{\otimes n} f
  = \tbcolvec{1}{\alpha}^{\otimes n} + i^r \tbcolvec{1}{-\alpha}^{\otimes n}.
 \]
 By Lemma~\ref{lem:vector:uniq}, there are two cases to consider, each of which has two more cases depending on $\varepsilon$.
 \begin{enumerate}
  \item Suppose $y a_0 - x b_0 = \alpha(x a_0 + y b_0)$, $y a_1 - x b_1 = -\alpha(x a_1 + y b_1)$, $(x a_0 + y b_0)^n = 1$, and $(x a_1 + y b_1)^n = i^r$.
  By rearranging the first two equations, we get
  \begin{equation} \label{eqn:a0b0_and_a1b1}
   (y - \alpha x) a_0 = (x + \alpha y) b_0
   \qquad \text{and} \qquad
   (y + \alpha x) a_1 = (x-\alpha y) b_1.
  \end{equation} 
  It cannot be the case that $a_0 = b_0 = 0$ or $y - \alpha x = x + \alpha y = 0$.
  If $a_0 = 0$, then $x + \alpha y = 0$, so $a_1 = -\sqrt{2} i b_1$ by~(\ref{eqn:a0b0_and_a1b1}) and $y \neq  0$ lest $x = 0$ as well.
  If $b_0 = 0$, then $y - \alpha x = 0$, so $\sqrt{2} i a_1 = b_1$, by the same argument.
  Now we consider the different cases for $\varepsilon$.
  \begin{enumerate}
   \item \label{case:A3_trans:+1}
   If $\varepsilon = 1$, then $a_1 = c (i a_0 - \sqrt{2} b_0)$ and $b_1 = c (\sqrt{2} a_0 + i b_0)$ by (\ref{eqn:a1b1-expression}).
   If $a_0 = 0$, then $a_1 = -c \sqrt{2} b_0$ and $b_1 = c i b_0$, which contradicts $a_1 = -\sqrt{2} i b_1$;
   if $b_0 = 0$, then $a_1 = c i a_0$ and $b_1 = c \sqrt{2} a_0$, which contradicts $\sqrt{2} i a_1 = b_1$.
   Thus, $(y - \alpha x) a_0 = (x + \alpha y) b_0 \neq 0$ by~(\ref{eqn:a0b0_and_a1b1}).
   Also from~(\ref{eqn:a0b0_and_a1b1}), $(y + \alpha x) a_1  = (x - \alpha y) b_1$.
   Then since $c \neq 0$ and using~(\ref{eqn:a1b1-expression}) with $\varepsilon = 1$, we get
   \[
    (y + \alpha x) \left(i a_0 - \sqrt{2} b_0\right) = (x - \alpha y) \left(\sqrt{2} a_0 + i b_0\right).
   \]
   Using $(y - \alpha x) a_0 = (x + \alpha y) b_0 \neq 0$, we get
   \[
    (y + \alpha x) \left(i (x + \alpha y) - \sqrt{2} (y - \alpha x)\right) = (x - \alpha y) \left(\sqrt{2} (x + \alpha y) + i (y - \alpha x)\right).
   \]
   This equation simplifies to $x^2 + y^2 = 0$, which is a contradiction.
   \item \label{case:A3_trans:-1}
   If $\varepsilon = -1$, then $a_1 = c (-i a_0 - \sqrt{2} b_0)$ and $b_1 = c (\sqrt{2} a_0 - i b_0)$, from  (\ref{eqn:a1b1-expression}).
   Then we get
   \begin{align*}
    x a_1 + y b_1
    &= x c \left(-i a_0 - \sqrt{2} b_0\right) + y c \left(\sqrt{2} a_0 - i b_0\right)\\
    &= c \left(-i (x a_0 + y b_0) + \sqrt{2}(y a_0 - x b_0)\right)\\
    &= c (x a_0 + y b_0),
   \end{align*}
   where in the third step, we used $y a_0 - x b_0 = \alpha(x a_0 + y b_0)$ from~(\ref{eqn:a0b0_and_a1b1}).
   Raising this equation to the $n$th power and using $(x a_0 + y b_0)^n = 1$ and $(x a_1 + y b_1)^n = i^r$,
   we conclude that $c^n = i^r$.
  \end{enumerate}
  \item Suppose $y a_0 - x b_0 = -\alpha(x a_0 + y b_0)$, $y a_1 - x b_1 = \alpha(x a_1 + y b_1)$, $(x a_0 + y b_0)^n = i^r$, and $(x a_1 + y b_1)^n = 1$.
  Now we consider the different cases for $\varepsilon$.
  \begin{enumerate}
   \item If $\varepsilon =  1$, 
   then $a_1 = c ( i a_0 - \sqrt{2} b_0)$ and $b_1 = c (\sqrt{2} a_0 + i b_0)$ by~(\ref{eqn:a1b1-expression}).
   Using similar reasoning to that in case~\ref{case:A3_trans:-1} leads to $(-c)^n i^r = 1$, so $c^n$ is a power of $i$.
   \item If $\varepsilon = -1$, then $a_1 = c (-i a_0 - \sqrt{2} b_0)$ and $b_1 = c (\sqrt{2} a_0 - i b_0)$ by~(\ref{eqn:a1b1-expression}).
   Using similar reasoning to that in case~\ref{case:A3_trans:+1} leads to a contradiction.
  \end{enumerate}
 \end{enumerate}

 For sufficiency, suppose the three equations hold for some $\varepsilon \in \{1,-1\}$ and some $r \in \{0,1,2,3\}$.
 Further assume $\varepsilon = 1$, in which case, the equations are
 \begin{equation} \label{eqn:theta:1/2}
  a_1 \left(\sqrt{2} a_0 + i b_0\right) = b_1 \left(i a_0 - \sqrt{2} b_0\right),
 \end{equation}
 as well as
 \begin{equation} \label{eqn:theta:1/2:2}
  a_1^n = i^r \left(i a_0 - \sqrt{2} b_0\right)^n
  \qquad \text{and} \qquad
  b_1^n = i^r \left(\sqrt{2} a_0 + i b_0\right)^n.
 \end{equation}
 From~(\ref{eqn:theta:1/2}), we have 
 \begin{equation} \label{a1-b1-expression-with-c}
  a_1 = c (i a_0 - \sqrt{2} b_0)
  \qquad \text{and} \qquad
  b_1 = c (\sqrt{2} a_0 + i b_0)
 \end{equation}
 for some $c \in \mathbb{C}$.
 In~(\ref{eqn:theta:1/2}), $a_1, b_1$ cannot be both zero.
 Similarly, $\sqrt{2} a_0 + i b_0, i a_0 - \sqrt{2} b_0$ cannot be both zero.
 Thus at least one equation in~(\ref{a1-b1-expression-with-c}) has both sides nonzero and we can always find some $c$ even if one factor is zero.
 We can write~(\ref{a1-b1-expression-with-c}) as
 \[
  \tbcolvec{a_1}{b_1} = c \tbmatrix{i}{-\sqrt{2}}{\sqrt{2}}{i} \tbcolvec{a_0}{b_0}.
 \]
 This implies that $a_0 a_1 + b_0 b_1 = c i (a_0^2 + b_0^2)$.
 Using~(\ref{eqn:theta:1/2:2}) or~(\ref{a1-b1-expression-with-c}),
 whichever equation is not zero on both sides,
 we have $c^n = i^r$.
 Since~(\ref{eqn:theta:1/2}) implies $\theta(f) = -\frac{1}{2}$,
 we know that $a_0^2 + b_0^2 \neq 0$ because otherwise $v_0$ is a multiple of $\tbcolvec{1}{\pm i}$,
 which makes $\theta(f) = -1$ regardless of $v_1$.
 
 We now define two orthogonal matrices $T_1 = \frac{1}{\sqrt{1 + i}} \tbmatrix{1}{\alpha}{-\alpha}{1}$ and $T_2 = \frac{1}{\sqrt{a_0^2 + b_0^2}} \tbmatrix{a_0}{b_0}{b_0}{-a_0}$.
 Also let $T = T_1 T_2 \in \mathbf{O}_2(\mathbb{C})$.
 For $f = \tbcolvec{a_0}{b_0}^{\otimes n} + \tbcolvec{a_1}{b_1}^{\otimes n}$,
 we want to calculate $T^{\otimes n} f$.
 First,
 \[
  T_2 \tbcolvec{a_0}{b_0} = \sqrt{a_0^2 + b_0^2} \tbcolvec{1}{0}
  \qquad \text{and} \qquad
  T \tbcolvec{a_0}{b_0} = \gamma \tbcolvec{1}{-\alpha},
 \]
 where $\gamma = \sqrt{\frac{a_0^2 + b_0^2}{1+i}}$.
 Furthermore, $a_1 b_0 - a_0 b_1 = \sqrt{2} i (a_0 a_1 + b_0 b_1) = - c \sqrt{2} (a_0^2 + b_0^2)$ by~\eqref{eqn:theta:1/2} and~\eqref{a1-b1-expression-with-c}.
 Then
 \[
  T_2 \tbcolvec{a_1}{b_1} = \frac{1}{\sqrt{a_0^2 + b_0^2}} \tbcolvec{a_0 a_1 + b_0 b_1}{a_1 b_0 - a_0 b_1} = c \sqrt{a_0^2 + b_0^2} \tbcolvec{i}{-\sqrt{2}}.
 \]
 It follows that
 \[
  T \tbcolvec{a_1}{b_1} = c \gamma  \tbmatrix{1}{\alpha}{-\alpha}{1} \tbcolvec{i}{-\sqrt{2}}
  =  c \gamma \tbcolvec{i - \sqrt{2} \alpha}{-i \alpha - \sqrt{2}}
  = -c \gamma \tbcolvec{1}{\alpha}.
 \]
 Thus
 \[
  T^{\otimes n} f = \gamma^n \left(\tbcolvec{1}{-\alpha}^{\otimes n} + (-c)^n \tbcolvec{1}{\alpha}^{\otimes n}\right).
 \]
 So $T$ transforms $f$ into the canonical form of $\mathscr{A}_3$.
 If we write out the orthogonal transformation $T$ explicitly,
 then $T = \tbmatrix{x}{y}{y}{-x}$ where
 \[
  x = \frac{a_0 + \alpha b_0}{\sqrt{(i + 1) \left(a_0^2 + b_0^2\right)}}
  \qquad \text{and} \qquad
  y=\frac{b_0 - \alpha a_0}{\sqrt{(i + 1) \left(a_0^2 + b_0^2\right)}}.
 \]

 When $\varepsilon = -1$, the argument is similar.
 In this case, $a_1 = c (-i a_0 - \sqrt{2} b_0)$ and $b_1 = c (\sqrt{2} a_0 - i b_0)$ for some $c \in \mathbb{C}$ satisfying $c^n = i^r$ and the entries of $T$ are
 \[
  x = \frac{a_0 - \alpha b_0}{\sqrt{(i + 1) \left(a_0^2 + b_0^2\right)}}
  \qquad \text{and} \qquad
  y = \frac{b_0 + \alpha a_0}{\sqrt{(i + 1) \left(a_0^2 + b_0^2\right)}}.
  \qedhere
 \]
\end{proof}

\begin{remark}
 Notice that either
 $a_1 (\sqrt{2} a_0 + i b_0) = b_1 ( i a_0 - \sqrt{2} b_0)$ or
 $a_1 (\sqrt{2} a_0 - i b_0) = b_1 (-i a_0 - \sqrt{2} b_0)$ implies $\theta(f) = -\frac{1}{2}$,
 unless $\det(\tbmatrix{a_0}{a_1}{b_0}{b_1}) = 0$.
\end{remark}

As mentioned before, $\mathscr{A}_2 = \mathscr{P}_2$ requires a stronger condition than just $\theta$.
If $f \in \mathscr{A}_2 = \mathscr{P}_2$, then $\theta(f) = -1$, but the reverse is not true.
If $f = v_0^{\otimes n} + v_1^{\otimes n}$ with $v_0 = [1,i]$ and $v_1$ is not a multiple of $[1,-i]$,
then $\theta(f) = -1$ but $f$ is not in $\mathscr{A}_2 = \mathscr{P}_2$,
since any orthogonal $H$ fixes $\{[1,i], [1,-i]\}$ set-wise,
up to a scalar multiple.

The next lemma, which appeared in~\cite{CGW16}, gives a characterization of $\mathscr{A}_2$.
It says that any signature in $\mathscr{A}_2$ is essentially in canonical form.
For completeness, we include its proof.

\begin{lemma}[\cite{CGW16}] \label{lem:single:P2}
 Let $f$ be a non-degenerate symmetric signature.
 Then $f \in \mathscr{A}_2$ iff $f$ is of the form
        $c \left(\tbcolvec{1}{ i}^{\otimes n}
         + \beta \tbcolvec{1}{-i}^{\otimes n}\right)$
 for some $c, \beta \neq 0$.
\end{lemma}

\begin{proof}
 Assume that $f = c \left(\tbcolvec{1}{ i}^{\otimes n}
                  + \beta \tbcolvec{1}{-i}^{\otimes n}\right)$ for some $c, \beta \ne 0$.
 Consider the orthogonal transformation $H = \tbmatrix{a}{b}{b}{-a}$,
 where $a = \frac{1}{2} \left(\beta^{\frac{1}{2n}} + \beta^{-\frac{1}{2n}}\right)$ and $b = \frac{1}{2i} \left(\beta^{\frac{1}{2n}} - \beta^{-\frac{1}{2n}}\right)$.
 We pick $a$ and $b$ in this way so that $a + b i = \beta^{\frac{1}{2n}}$, $a - b i = \beta^{-\frac{1}{2n}}$, and $(a + b i) (a - b i) = a^2 + b^2 = 1$.
 Also $\left(\frac{a + b i}{a - b i}\right)^n = \beta$.
 Then
 \begin{align*}
  H^{\otimes n} f
  &= c \left(\tbcolvec{a + b i}{-a i + b}^{\otimes n} + \beta \tbcolvec{a - b i}{a i + b}^{\otimes n}\right)\\
  &= c \left((a + b i)^n \tbcolvec{1}{-i}^{\otimes n} + (a - b i)^n \beta \tbcolvec{1}{i}^{\otimes n}\right)\\
  &= c \sqrt{\beta} \left(\tbcolvec{1}{-i}^{\otimes n} + \tbcolvec{1}{i}^{\otimes n}\right),
 \end{align*}
 so $f$ can be written as
 \[
  f = c \sqrt{\beta} (H^{-1})^{\otimes n} \left(\tbcolvec{1}{-i}^{\otimes n} + \tbcolvec{1}{i}^{\otimes n}\right).
 \]
 Therefore $f \in \mathscr{A}_2$.
 
 On the other hand, the desired form $f = c (\tbcolvec{1}{i}^{\otimes n} + \beta \tbcolvec{1}{i}^{\otimes n})$
 follows from the fact that $\{\tbcolvec{1}{i}, \tbcolvec{1}{-i}\}$ is fixed setwise under any orthogonal transformation up to nonzero constants.
\end{proof}

\begin{remark}
 Notice that $\theta(v_0, v_1) = -1$ for linearly independent $v_0$ and $v_1$
 if and only if at least one of $v_0, v_1$ is $\tbcolvec{1}{i}$ or $\tbcolvec{1}{- i}$, up to a nonzero scalar.
\end{remark}

We now present the polynomial-time algorithm to check if $f \in \mathscr{A}_1 \union \mathscr{A}_2 \union \mathscr{A}_3$.

\begin{lemma} \label{lem:single:affine}
 Given a non-degenerate symmetric signature $f$ of arity at least~$3$,
 there is a polynomial-time algorithm to decide whether $f \in \mathscr{A}_k$ for each $k \in \{1,2,3\}$.
 If so, $k$ is unique and at least one corresponding orthogonal transformation can be found in polynomial time.
\end{lemma}

\begin{proof}
 First we check if $f$ satisfies a second order recurrence relation.
 If it does, then the coefficients $(a,b,c)$ of the second order recurrence relation are unique up to a nonzero scalar by Lemma~\ref{lem:symmetric:2nd_ord_rec_rel:unique}.
 If the coefficients satisfy $b^2 - 4 a c \ne 0$,
 then by Lemma~\ref{lem:symmetric:2nd_ord_rec_rel:characterization},
 we can express $f$ as $v_0^{\otimes n} + v_1^{\otimes n}$,
 where $v_0$ and $v_1$ are linearly independent and $\arity(f) = n$.
 All of this must be true for $f$ to be in $\mathscr{A}_1 \union \mathscr{A}_2 \union \mathscr{A}_3$.
 With this alternate expression for $f$,
 we apply Lemma~\ref{lem:single:A1}, Lemma~\ref{lem:single:P2}, and Lemma~\ref{lem:single:A3} to decide if $f \in \mathscr{A}_k$ for each $k \in \{1,2,3\}$ respectively.
 These sets are disjoint by Lemma~\ref{lem:symmetric:angle:AP-trans},
 so there can be at most one $k$ such that $f \in \mathscr{A}_k$.
\end{proof}

%% file: 6.2.Symmetric.Affine.Set.tex
\subsection{Set of Symmetric Signatures}

We first show that if a non-degenerate signature $f$ of arity at least~$3$ is in $\mathscr{A}_1$ or $\mathscr{A}_3$,
then for any set $\mathcal{F}$ containing $f$,
there are only a small constant number of transformations to check to decide whether $\mathcal{F}$ is $\mathscr{A}$-transformable.
If $f \in \mathscr{A}_2$,
then there can be more than a constant number of transformations to check.
However, this number is at most linear in the arity of $f$.

Notice that any non-degenerate symmetric signature $f \in \mathscr{A}$ of arity at least~$3$ is in $\mathscr{F}_{123}$ (introduced in Section \ref{sec:prelim:tractable}),
which contains signatures expressed as a sum of two tensor powers.
Therefore $\theta(f)$ is well-defined.
By Lemma~\ref{lem:affine:trans}, to check $\mathscr{A}$-transformability,
we may restrict our attention to the sets $\mathscr{A}$ and $\tbmatrix{1}{0}{0}{\alpha} \mathscr{A}$ up to orthogonal transformations.
In particular,
\begin{equation} \label{eqn:theta:symmetric}
 \theta(f)
 =
 \begin{cases}
             0 & \text{if } f \in \mathscr{F}_1 \union \mathscr{F}_2 \union \tbmatrix{1}{0}{0}{\alpha} \mathscr{F}_1,\\
            -1 & \text{if } f \in \mathscr{F}_3,\\
  -\frac{1}{2} & \text{if } f \in \tbmatrix{1}{0}{0}{\alpha} (\mathscr{F}_2 \union \mathscr{F}_3).
 \end{cases}
\end{equation}

\begin{lemma} \label{lem:trans:A1}
 Let $\mathcal{F}$ be a set of symmetric signatures and suppose $\mathcal{F}$ contains
 a non-degenerate signature $f \in \mathscr{A}_1$ of arity $n \ge 3$ with $H \in \mathbf{O}_2(\mathbb{C})$.
 Then $\mathcal{F}$ is $\mathscr{A}$-transformable
 iff $\mathcal{F}$ is a subset of
 $H \mathscr{A}$, or
 $H \tbmatrix{1}{1}{1}{-1} \mathscr{A}$, or
 $H \tbmatrix{1}{1}{1}{-1} \tbmatrix{1}{0}{0}{\alpha} \mathscr{A}$.
\end{lemma}

\begin{proof}
 Sufficiency follows from Lemma~\ref{lem:affine:trans} and both $H, H_2 = \frac{1}{\sqrt{2}} \tbmatrix{1}{1}{1}{-1} \in \mathbf{O_2}(\mathbb{C})$.
  
 Before we prove necessity, we first claim that without loss of generality,
 we may assume $H \in \mathbf{O}_2(\mathbb{C}) \setminus \mathbf{SO}_2(\mathbb{C})$.
 If $H \in \mathbf{SO}_2(\mathbb{C})$,
 we let $\widetilde{H} = H \tbmatrix{0}{1}{1}{0} \in \mathbf{O}_2(\mathbb{C}) \setminus \mathbf{SO}_2(\mathbb{C})$.
 Then $f \in \mathscr{A}_1$ also with $\widetilde{H}$.
 From $\tbmatrix{0}{1}{1}{0} \in \StabA$, it follows that $\widetilde{H} \mathscr{A} = H \mathscr{A}$. 
 Also
 $\tbmatrix{0}{1}{1}{0} \tbmatrix{1}{1}{1}{-1} \tbmatrix{1}{0}{0}{\alpha}
 = \tbmatrix{1}{-1}{1}{1} \tbmatrix{1}{0}{0}{\alpha}
 = \tbmatrix{1}{1}{1}{-1} \tbmatrix{1}{0}{0}{-1} \tbmatrix{1}{0}{0}{\alpha}
 = \tbmatrix{1}{1}{1}{-1} \tbmatrix{1}{0}{0}{\alpha} \tbmatrix{1}{0}{0}{-1}$,
 and $\tbmatrix{1}{0}{0}{-1} \in \StabA$. 
 It follows that
 $\widetilde{H} \tbmatrix{1}{1}{1}{-1} \tbmatrix{1}{0}{0}{\alpha} \mathscr{A}
 = H \tbmatrix{1}{1}{1}{-1} \tbmatrix{1}{0}{0}{\alpha} \mathscr{A}$. 

 Suppose $\mathcal{F}$ is $\mathscr{A}$-transformable.
 By Lemma~\ref{lem:affine:trans:so2},
 there exists an $H' \in \mathbf{SO}_2(\mathbb{C})$ such that $\mathcal{F} \subseteq H' \mathscr{A}$
 or $\mathcal{F} \subseteq H' \tbmatrix{1}{0}{0}{\alpha} \mathscr{A}$.
 We only need to show there exists an $M \in \StabA$,
 such that $H' = H M$
 in the first case,
 and in the second case $H' = H \tbmatrix{1}{1}{1}{-1} M$,
 and $M \tbmatrix{1}{0}{0}{\alpha} = \tbmatrix{1}{0}{0}{\alpha} M'$ for some $M' \in \StabA$.

 Since $f \in \mathscr{A}_1$ with $H$,
 after a suitable normalization by a nonzero scalar,
 we have
 \[
  f = H^{\otimes n} \left(\tbcolvec{1}{ 1}^{\otimes n}
                  + \beta \tbcolvec{1}{-1}^{\otimes n}\right),
 \]
 where $\beta = \alpha^{tn+2r}$ for some $r \in \{0,1,2,3\}$ and $t \in \{0,1\}$.
 Let $g = (H'^{-1})^{\otimes n} f$ and $T = H'^{-1} H$ so that
 \[
  g = T^{\otimes n} \left(\tbcolvec{1}{ 1}^{\otimes n}
                            + \beta \tbcolvec{1}{-1}^{\otimes n}\right).
 \]
 Note that $T \in \mathbf{O}_2(\mathbb{C}) \setminus \mathbf{SO}_2(\mathbb{C})$
 since $H' \in \mathbf{SO}_2(\mathbb{C})$ and $H \in \mathbf{O}_2(\mathbb{C}) \setminus \mathbf{SO}_2(\mathbb{C})$.
 Thus $T = T^{-1}$ and $H T = H'$.
 Let $T = \tbmatrix{a}{b}{b}{-a}$ for some $a, b \in \mathbb{C}$ such that $a^2 + b^2 = 1$.
 There are two possibilities according to whether $\mathcal{F} \subseteq H' \mathscr{A}$ or $\mathcal{F} \subseteq H' \tbmatrix{1}{0}{0}{\alpha} \mathscr{A}$.
 \begin{enumerate}
  \item If $\mathcal{F} \subseteq H' \mathscr{A}$, then $g \in \mathscr{F}_{123}$ since $g$ is symmetric and non-degenerate.
  Since $\theta(g) = 0$, by~\eqref{eqn:theta:symmetric}, $g \in \mathscr{F}_1$ or $g \in \mathscr{F}_2$.
  We discuss the two cases of $g$ separately.
  \begin{itemize}
   \item Suppose $g \in \mathscr{F}_1$.
   Then we have
   \[
    T^{\otimes n} \left(\tbcolvec{1}{ 1}^{\otimes n}
                + \beta \tbcolvec{1}{-1}^{\otimes n}\right)
    = \lambda \left(\tbcolvec{1}{0}^{\otimes n}
              + i^t \tbcolvec{0}{1}^{\otimes n}\right)
   \]
   for some $\lambda \neq 0$ and $t \in \{0,1,2,3\}$.
   Plugging in the expression for $T$, we have
   \[
      \left(\tbcolvec{a+b}{b-a}^{\otimes n}
    + \beta \tbcolvec{a-b}{a+b}^{\otimes n}\right)
    = \lambda \left(\tbcolvec{1}{0}^{\otimes n}
              + i^t \tbcolvec{0}{1}^{\otimes n}\right).
   \]
   Then by Lemma~\ref{lem:vector:uniq}, we have $a+b=0$ or $a-b=0$.
   Together with $a^2 + b^2 = 1$,
   we can solve for $T = \frac{1}{\sqrt{2}}\tbmatrix{1}{1}{1}{-1}$
   or $T = \frac{1}{\sqrt{2}} \tbmatrix{1}{-1}{-1}{-1} = \frac{1}{\sqrt{2}} \tbmatrix{1}{1}{1}{-1} \tbmatrix{0}{-1}{1}{0}$,
   up to a constant multiple $\pm 1$.
   Since $\tbmatrix{0}{-1}{1}{0} \in \StabA$, we have $T \in \StabA$, so we are done.
   \item Suppose $g \in \mathscr{F}_2$.
   Then we have
   \[
    T^{\otimes n} \left(\tbcolvec{1}{ 1}^{\otimes n}
                + \beta \tbcolvec{1}{-1}^{\otimes n}\right)
    = \lambda \left(\tbcolvec{1}{ 1}^{\otimes n}
              + i^t \tbcolvec{1}{-1}^{\otimes n}\right)
   \]
   for some $\lambda \neq 0$ and $t \in \{0,1,2,3\}$.
   Plugging in the expression for $T$, we have
   \[
      \left(\tbcolvec{a+b}{b-a}^{\otimes n}
    + \beta \tbcolvec{a-b}{a+b}^{\otimes n}\right)
    = \lambda \left(\tbcolvec{1}{ 1}^{\otimes n}
              + i^t \tbcolvec{1}{-1}^{\otimes n}\right)
   \]
   Then by Lemma~\ref{lem:vector:uniq}, we have $a+b=a-b$ or $a+b=-(a-b)$.
   Therefore either $a=0$ or $b=0$.
   Thus $T = \pm \tbmatrix{1}{0}{0}{-1}$ or $T= \pm \tbmatrix{0}{1}{1}{0}$ and both matrices are in $\StabA$.
  \end{itemize}
  \item If $\mathcal{F} \subseteq H' \tbmatrix{1}{0}{0}{\alpha} \mathscr{A}$,
  then we have $g \in \tbmatrix{1}{0}{0}{\alpha} \mathscr{F}_{123}$.
  Since $\theta(g) = 0$, by~\eqref{eqn:theta:symmetric}, $g \in \tbmatrix{1}{0}{0}{\alpha} \mathscr{F}_1$.
  That is,
  \[
   T^{\otimes n} \left(\tbcolvec{1}{ 1}^{\otimes n}
               + \beta \tbcolvec{1}{-1}^{\otimes n}\right)
   = \lambda \tbmatrix{1}{0}{0}{\alpha}^{\otimes n} \left(\tbcolvec{1}{0}^{\otimes n}
                                                    + i^t \tbcolvec{0}{1}^{\otimes n}\right)
   = \lambda \left(\tbcolvec{1}{0}^{\otimes n}
    + i^t \alpha^n \tbcolvec{0}{1}^{\otimes n}\right)
  \]
  for some $\lambda \neq 0$.
  This is essentially the same as the case where $g \in \mathscr{F}_1$ above,
  except that the coefficients are different.
  However, the coefficients do not affect the argument and our conclusion in this case that $T = \frac{1}{\sqrt{2}} \tbmatrix{1}{1}{1}{-1}$
  or $T = \frac{1}{\sqrt{2}} \tbmatrix{1}{1}{1}{-1} \tbmatrix{0}{-1}{1}{0}$,
  up to a constant multiple $\pm 1$.
  Notice that $\tbmatrix{0}{-1}{1}{0} \in \StabA$.
  Moreover, 
  \begin{align*}
   \tbmatrix{0}{-1}{1}{0} \tbmatrix{1}{0}{0}{\alpha}
   &= \tbmatrix{0}{-\alpha}{1}{0} \\
   &= \tbmatrix{1}{0}{0}{\alpha} \tbmatrix{0}{-\alpha}{\alpha^{-1}}{0} \\
   &= -\alpha \tbmatrix{1}{0}{0}{\alpha} \tbmatrix{0}{1}{i}{0},
  \end{align*}
  and $\tbmatrix{0}{1}{i}{0} \in \StabA$.
  \qedhere
 \end{enumerate} 
\end{proof}

\begin{lemma} \label{lem:trans:A2}
 Let $\mathcal{F}$ be a set of symmetric signatures and suppose $\mathcal{F}$ contains
 a non-degenerate signature $f \in \mathscr{A}_2$ of arity $n \ge 3$.
 Then there exists a set $\mathcal{H} \subseteq \mathbf{O}_2(\mathbb{C})$ of size $O(n)$ such that $\mathcal{F}$ is $\mathscr{A}$-transformable
 iff there exists an $H \in \mathcal{H}$ such that $\mathcal{F} \subseteq H \mathscr{A}$.
 Moreover $\mathcal{H}$ can be computed in polynomial time in the input length of the symmetric signature $f$.
\end{lemma}

\begin{proof}
 Sufficiency is trivial by Lemma~\ref{lem:affine:trans:so2}.

 Suppose $\mathcal{F}$ is $\mathscr{A}$-transformable.
 By Lemma~\ref{lem:affine:trans:so2}, 
 there exists an $H \in \mathbf{SO}_2(\mathbb{C})$ such that $\mathcal{F} \subseteq H \mathscr{A}$
 or $\mathcal{F} \subseteq H \tbmatrix{1}{0}{0}{\alpha} \mathscr{A}$.
 In the first case, we show that the number of choices of $H$ can be limited to $O(n)$.
 Then we show that the second case is impossible.

 Since $f \in \mathscr{A}_2$,
 after a suitable normalization by a nonzero scalar,
 we have
 \[
  f = \tbcolvec{1}{i}^{\otimes n} + \nu \tbcolvec{1}{-i}^{\otimes n}
 \]
 for some $\nu \neq 0$ by Lemma~\ref{lem:single:P2}.
 Let $g = (H^{-1})^{\otimes n} f$.
 Then
 \[
  g = (H^{-1})^{\otimes n} \left(\tbcolvec{1}{ i}^{\otimes n}
                           + \nu \tbcolvec{1}{-i}^{\otimes n}\right).
 \]
 There are two possibilities according to whether $\mathcal{F} \subseteq H \mathscr{A}$ or $\mathcal{F} \subseteq H \tbmatrix{1}{0}{0}{\alpha} \mathscr{A}$.

 \begin{enumerate}
  \item Suppose $\mathcal{F} \subseteq H \mathscr{A}$.
  Therefore $g \in \mathscr{F}_{123}$.
  Since $\theta(g) = -1$, by~\eqref{eqn:theta:symmetric}, $g \in \mathscr{F}_3$.
  Then we have
  \[
   (H^{-1})^{\otimes n} \left(\tbcolvec{1}{ i}^{\otimes n}
                        + \nu \tbcolvec{1}{-i}^{\otimes n}\right)
   = \lambda \left(\tbcolvec{1}{ i}^{\otimes n}
             + i^r \tbcolvec{1}{-i}^{\otimes n}\right)
  \]
  for some $\lambda \neq 0$ and $r \in \{0,1,2,3\}$.
  Because $H^{-1} \in \mathbf{SO}_2(\mathbb{C})$, 
  we may assume that $H^{-1}$ is of the form $\tbmatrix{a}{b}{-b}{a}$	where $a^2 + b^2 = 1$.
  Therefore
  \begin{align*}
   \lambda \left(\tbcolvec{1}{ i}^{\otimes n}
           + i^r \tbcolvec{1}{-i}^{\otimes n}\right)
   &= \tbmatrix{a}{b}{-b}{a}^{\otimes n} \left(\tbcolvec{1}{ i}^{\otimes n}
                                         + \nu \tbcolvec{1}{-i}^{\otimes n}\right)\\
   &=    (a+bi)^n \tbcolvec{1}{ i}^{\otimes n}
   + \nu (a-bi)^n \tbcolvec{1}{-i}^{\otimes n}.
  \end{align*}
  Comparing the coefficients, by Lemma~\ref{lem:vector:uniq}, we have
  \[
   \lambda = (a+bi)^n
   \qquad \text{and} \qquad
   \lambda i^r = \nu (a-bi)^n.
  \]
  Hence,
  \[
   i^r (a+bi)^n = \nu (a-bi)^n.
  \]
  Since $(a+bi) (a-bi) = a^2 + b^2 = 1$, we know that $(a+bi)^{2n} = \nu i^{-r}$.
  Therefore $a + b i = \omega_{2n} (\nu i^{-r})^{1/2n}$,
  where $\omega_{2n}$ is a $2n$-th root of unity.
  There are~$4$ choices for $r$, and $2n$ choices for $\omega_{2n}$.
  However, $a - b i = \frac{1}{a+bi}$, and $(a,b)$ can be solved from $(a+bi,a-bi)$.
  Hence there are only $O(n)$ choices for $H$, depending on $f$.

  \item Suppose $\mathcal{F} \subseteq H \tbmatrix{1}{0}{0}{\alpha} \mathscr{A}$.
  Then $g \in \tbmatrix{1}{0}{0}{\alpha} \mathscr{F}_{123}$.
  However, $\theta(g) = -1$,
  which contradicts~\eqref{eqn:theta:symmetric}.
  \qedhere
 \end{enumerate}
\end{proof}

\begin{lemma} \label{lem:trans:A3}
 Let $\mathcal{F}$ be a set of symmetric signatures and suppose $\mathcal{F}$ contains
 a non-degenerate signature $f \in \mathscr{A}_3$ of arity $n \ge 3$ with $H \in \mathbf{O}_2(\mathbb{C})$.
 Then $\mathcal{F}$ is $\mathscr{A}$-transformable iff $\mathcal{F} \subseteq H \tbmatrix{1}{0}{0}{\alpha} \mathscr{A}$.
\end{lemma}

\begin{proof}
 Sufficiency is trivial by Lemma~\ref{lem:affine:trans:so2}. 

 Suppose $\mathcal{F}$ is $\mathscr{A}$-transformable.
 As in the proof of Lemma~\ref{lem:trans:A1},
 we may assume that $H \in \mathbf{O}_2(\mathbb{C}) \setminus \mathbf{SO}_2(\mathbb{C})$.
 By Lemma~\ref{lem:affine:trans:so2},
 there exists an $H' \in \mathbf{SO}_2(\mathbb{C})$ such that $\mathcal{F} \subseteq H'\mathscr{A}$
 or $\mathcal{F} \subseteq H' \tbmatrix{1}{0}{0}{\alpha} \mathscr{A}$.
 We show the first case is impossible.
 Then in the second case,
 we show that there exists an $M$ such that $H' = H M$,
 where $M \tbmatrix{1}{0}{0}{\alpha} = \tbmatrix{1}{0}{0}{\alpha} M'$ for some $M' \in \StabA$.

 Since $f \in \mathscr{A}_3$ with $H$,
 after a suitable normalization by a nonzero scalar,
 we have
 \[
  f = H^{\otimes n} \left(\tbcolvec{1}{ \alpha}^{\otimes n}
                    + i^r \tbcolvec{1}{-\alpha}^{\otimes n}\right)
 \]
 for some $r \in \{0,1,2,3\}$.
 Let $g = (H'^{-1})^{\otimes n} f$ and $T = H'^{-1} H$ so that
 \[
  g = T^{\otimes n} \left(\tbcolvec{1}{ \alpha}^{\otimes n}
                              + i^r \tbcolvec{1}{-\alpha}^{\otimes n}\right).
 \]
 Note that $T \in \mathbf{O}_2(\mathbb{C}) \setminus \mathbf{SO}_2(\mathbb{C})$
 since $H' \in \mathbf{SO}_2(\mathbb{C})$ and $H \in \mathbf{O}_2(\mathbb{C}) \setminus \mathbf{SO}_2(\mathbb{C})$.
 Thus $T = T^{-1}$ and $H T = H'$.
 Let $T = \tbmatrix{a}{b}{b}{-a}$ for some $a, b \in \mathbb{C}$ such that $a^2 + b^2 = 1$.
 There are two possibilities according to whether $\mathcal{F} \subseteq H' \mathscr{A}$ or $\mathcal{F} \subseteq H' \tbmatrix{1}{0}{0}{\alpha} \mathscr{A}$.
 \begin{enumerate}
  \item Suppose $\mathcal{F} \subseteq H'\mathscr{A}$.
  Then $g = (H'^{-1})^{\otimes n} f \in \mathscr{F}_{123}$.
  However, $\theta(g) = -\frac{1}{2}$,
  which contradicts \eqref{eqn:theta:symmetric}.
  \item Suppose $\mathcal{F} \subseteq H' \tbmatrix{1}{0}{0}{\alpha} \mathscr{A}$.
  Then $g \in \tbmatrix{1}{0}{0}{\alpha} \mathscr{F}_{123}$,
  so $\theta(g) = -\frac{1}{2}$ and $g \in \tbmatrix{1}{0}{0}{\alpha} (\mathscr{F}_2 \union \mathscr{F}_3)$ by~\eqref{eqn:theta:symmetric}.
  We discuss the these two cases separately.
  \begin{itemize}
   \item Suppose $g\in\tbmatrix{1}{0}{0}{\alpha}\mathscr{F}_2$.
   Then we have
   \begin{align*}
    T^{\otimes n} \left(\tbcolvec{1}{ \alpha}^{\otimes n}
                  + i^r \tbcolvec{1}{-\alpha}^{\otimes n}\right)
    &= \lambda \tbmatrix{1}{0}{0}{\alpha}^{\otimes n} \left(\tbcolvec{1}{ 1}^{\otimes n}
                                                      + i^t \tbcolvec{1}{-1}^{\otimes n}\right)\\
    &= \lambda \left(\tbcolvec{1}{ \alpha}^{\otimes n}
               + i^t \tbcolvec{1}{-\alpha}^{\otimes n}\right)
   \end{align*}
   for some $\lambda \neq 0$ and $t \in \{0,1,2,3\}$.
   Plugging in the expression for $T$,
   we have
   \[
    \left(\tbcolvec{a + \alpha b}{b - \alpha a}^{\otimes n}
    + i^r \tbcolvec{a - \alpha b}{b + \alpha a}^{\otimes n}\right)
    = \lambda \left(\tbcolvec{1}{ \alpha}^{\otimes n}
              + i^t \tbcolvec{1}{-\alpha}^{\otimes n}\right).
   \]
   Then by Lemma~\ref{lem:vector:uniq},
   we have either
   \[
    b - a \alpha = \alpha (a + b \alpha)
    \qquad \text{and} \qquad
    b + a \alpha = -\alpha (a - b \alpha)
   \]
   or
   \[
    b - a \alpha = -\alpha (a + b \alpha)
    \qquad \text{and} \qquad
    b + a \alpha = \alpha (a - b \alpha).
   \]
   The first case is impossible.
   In the second case,
   we have $a = \pm 1$ and $b=0$.
   This implies $T = \pm \tbmatrix{1}{0}{0}{-1} \in \StabA$, which commutes with $\tbmatrix{1}{0}{0}{\alpha}$.
   \item Suppose $g \in \tbmatrix{1}{0}{0}{\alpha} \mathscr{F}_3$.
   Then we have
   \begin{align*}
    T^{\otimes n} \left(\tbcolvec{1}{ \alpha}^{\otimes n}
                  + i^r \tbcolvec{1}{-\alpha}^{\otimes n}\right)
    &= \lambda \tbmatrix{1}{0}{0}{\alpha}^{\otimes n} \left(\tbcolvec{1}{ i}^{\otimes n}
                                                      + i^t \tbcolvec{1}{-i}^{\otimes n}\right)\\
    &= \lambda \left(\tbcolvec{1}{ \alpha i}^{\otimes n}
               + i^t \tbcolvec{1}{-\alpha i}^{\otimes n}\right)
   \end{align*}
   for some $\lambda \neq 0$ and $t \in \{0,1,2,3\}$.
   Plugging in the expression for $T$,
   we have
   \[
    \left(\tbcolvec{a + \alpha b}{b - \alpha a}^{\otimes n}
    + i^r \tbcolvec{a - \alpha b}{b + \alpha a}^{\otimes n}\right)
    = \lambda \left(\tbcolvec{1}{ \alpha i}^{\otimes n}
              + i^t \tbcolvec{1}{-\alpha i}^{\otimes n}\right).
   \]
   Then by Lemma~\ref{lem:vector:uniq},
   we have either
   \[
    b - a \alpha = \alpha i (a + b \alpha)
    \qquad \text{and} \qquad
    b + a \alpha = -\alpha i (a - b \alpha)
   \]
   or
   \[
    b - a \alpha = -\alpha i (a + b \alpha)
    \qquad \text{and} \qquad
    b + a \alpha = \alpha i (a - b \alpha).
   \]
   The first case is impossible.
   In the second case,
   we have $a = 0$ and $b = \pm 1$.
   This implies that $T = \pm \tbmatrix{0}{1}{1}{0}$.
   Note that $\tbmatrix{0}{1}{1}{0} \tbmatrix{1}{0}{0}{\alpha} = \tbmatrix{1}{0}{0}{\alpha} \tbmatrix{0}{\alpha}{\alpha^{-1}}{0}$
   and $\tbmatrix{0}{\alpha}{\alpha^{-1}}{0} = \alpha^{-1} \tbmatrix{0}{i}{1}{0} \in\StabA$.
   \qedhere
  \end{itemize}
 \end{enumerate}
\end{proof}

Now we are ready to show how to decide if a finite set of signatures is $\mathscr{A}$-transformable.
To avoid trivialities,
we assume $\mathcal{F}$ contains a non-degenerate signature of arity at least~$3$.
If every non-degenerate signature in $\mathcal{F}$ has arity at most two,
then $\Holant(\mathcal{F})$ is tractable.

\begin{theorem}
 There is a polynomial-time algorithm to decide,
 for any finite input set $\mathcal{F}$ of symmetric signatures containing a non-degenerate signature $f$ of arity $n \geq 3$,
 whether $\mathcal{F}$ is $\mathscr{A}$-transformable.
\end{theorem}

\begin{proof}
 By Lemma~\ref{lem:single:affine},
 we can decide if $f$ is in $\mathscr{A}_k$ for some $k \in \{1,2,3\}$.
 If not, then by Lemma~\ref{lem:cha:affine}, $\mathcal{F}$ is not $\mathscr{A}$-transformable.
 Otherwise, $f \in \mathscr{A}_k$ for some unique $k$.
 Depending on $k$,
 we apply Lemma~\ref{lem:trans:A1}, Lemma~\ref{lem:trans:A2}, or Lemma~\ref{lem:trans:A3} to check if $\mathcal{F}$ is $\mathscr{A}$-transformable.
\end{proof}

%% file: 7.Symmetric.Product.tex
\section{Symmetric \texorpdfstring{$\mathscr{P}$}{P}-transformable Signatures} \label{sec:symmetric:product}

To decide if a signature set is $\mathscr{P}$-transformable,
we face the same issue as in the $\mathscr{A}$-transformable case.
Namely, a symmetric signature of arity $n$ is given by $n + 1$ values, instead of $2^n$ values.
This exponentially more succinct representation requires us to find a more efficient algorithm.

The next lemma tells us how to decide membership in $\mathscr{P}_1$ for signatures of arity at least~$3$.

\begin{lemma} \label{lem:single:P1}
 Let $f = v_0^{\otimes n} + v_1^{\otimes n}$ be a symmetric signature of arity $n \geq 3$,
 where $v_0$ and $v_1$ are linearly independent.
 Then $f \in \mathscr{P}_1$ iff $\theta(f) = 0$.
\end{lemma}

\begin{proof}
 Necessity is clear by Lemma~\ref{lem:symmetric:angle:AP-trans} and sufficiency follows from Lemma~\ref{lem:single:theta=0:H-less_form}.
\end{proof}

Since $\mathscr{A}_2 = \mathscr{P}_2$,
the membership problem for $\mathscr{P}_2$ is handled by Lemma~\ref{lem:single:P2}.
Using Lemma~\ref{lem:single:P1} and Lemma~\ref{lem:single:P2},
we can efficiently decide membership in $\mathscr{P}_1 \union \mathscr{P}_2$.

\begin{lemma} \label{lem:single:product}
 Given a non-degenerate symmetric signature $f$ of arity at least $3$, 
 there is a polynomial-time algorithm to decide whether $f \in \mathscr{P}_k$ for some $k \in \{1,2\}$.
 If so, $k$ is unique and at least one corresponding orthogonal transformation can be found in polynomial time.
\end{lemma}

\begin{proof}
 First we check if $f$ satisfies a second order recurrence relation.
 If it does, then the coefficients $(a,b,c)$ of the second order recurrence relation are unique up to a nonzero scalar by Lemma~\ref{lem:symmetric:2nd_ord_rec_rel:unique}.
 If the coefficients satisfy $b^2 - 4 a c \ne 0$,
 then by Lemma~\ref{lem:symmetric:2nd_ord_rec_rel:characterization},
 we can express $f$ as $v_0^{\otimes n} + v_1^{\otimes n}$,
 where $v_0$ and $v_1$ are linearly independent and $\arity(f) = n$.
 All of this must be true for $f$ to be in $\mathscr{P}_1 \union \mathscr{P}_2$.
 With this alternate expression for $f$,
 we apply Lemma~\ref{lem:single:P1} and Lemma~\ref{lem:single:P2} to decide if $f \in \mathscr{P}_k$ for some $k \in \{1,2\}$ respectively.
 These sets are disjoint by Lemma~\ref{lem:symmetric:angle:AP-trans},
 so there can be at most one $k$ such that $f \in \mathscr{P}_k$.
\end{proof}

Like the symmetric affine case,
the following lemmas assume the signature set $\mathcal{F}$ contains a non-degenerate signature of arity at least~$3$ in $\mathscr{P}_1$ or $\mathscr{P}_2$.
Unlike the symmetric affine case,
the number of transformations to be checked to decide whether $\mathcal{F}$ is $\mathscr{P}$-transformable is always a small constant.

\begin{lemma} \label{lem:trans:P1}
 Let $\mathcal{F}$ be a set of symmetric signatures and suppose $\mathcal{F}$ contains
 a non-degenerate signature $f \in \mathscr{P}_1$ of arity $n \ge 3$ with $H \in \mathbf{O}_2(\mathbb{C})$.
 Then $\mathcal{F}$ is $\mathscr{P}$-transformable iff $\mathcal{F} \subseteq H \tbmatrix{1}{1}{1}{-1} \mathscr{P}$.
\end{lemma}

\begin{proof}
 Sufficiency is trivial by Lemma~\ref{lem:product:trans}. 
 
 Suppose $\mathcal{F}$ is $\mathscr{P}$-transformable.
 As in the proof of Lemma~\ref{lem:trans:A1}, we may assume $H \in \mathbf{O}_2(\mathbb{C}) \setminus \mathbf{SO}_2(\mathbb{C})$.
 Then by Lemma~\ref{lem:product:trans:so2},
 there exists an $H' \in \mathbf{SO}_2(\mathbb{C})$ such that
 $\mathcal{F} \subseteq H' \mathscr{P}$ or $\mathcal{F}\subseteq H' \tbmatrix{1}{1}{i}{-i} \mathscr{P}$,
 where in the second case we can take $H' = I_2$.
 In the first case, we show that there exists an $M \in \StabP$ such that $H' = H \tbmatrix{1}{1}{1}{-1} M$.
 Then we show that the second case is impossible.

 Since $f \in \mathscr{P}_1$ with $H$,
 after a suitable normalization by a nonzero scalar,
 we have
 \[
  f = H^{\otimes n} \left(\tbcolvec{1}{ 1}^{\otimes n}
                  + \beta \tbcolvec{1}{-1}^{\otimes n}\right)
 \]
 for some $\beta \neq 0$.
 Let $g = (H'^{-1})^{\otimes n} f$ and $T = H'^{-1} H$ so that
 \[
  g = T^{\otimes n} \left(\tbcolvec{1}{ 1}^{\otimes n}
                            + \beta \tbcolvec{1}{-1}^{\otimes n}\right).
 \]
 Note that $T \in \mathbf{O}_2(\mathbb{C}) \setminus \mathbf{SO}_2(\mathbb{C})$
 since $H' \in \mathbf{SO}_2(\mathbb{C})$ and $H \in \mathbf{O}_2(\mathbb{C}) \setminus \mathbf{SO}_2(\mathbb{C})$.
 Thus $T = T^{-1}$ and $H T = H'$.
 \begin{enumerate}
  \item Suppose $\mathcal{F} \subseteq H' \mathscr{P}$.
  Then $g$ must be a generalized equality since $g \in \mathscr{P}$ with arity $n \ge 3$.
  The only symmetric non-degenerate generalized equalities in $\mathscr{P}$ with arity $n \ge 3$ have the form
  $\lambda \left(\tbcolvec{1}{0}^{\otimes n}
        + \beta' \tbcolvec{0}{1}^{\otimes n}\right)$, for some $\lambda, \beta' \neq 0$.
  Thus
  \[
   T^{\otimes n} \left(\tbcolvec{1}{ 1}^{\otimes n}
               + \beta \tbcolvec{1}{-1}^{\otimes n}\right)
   = \lambda \left(\tbcolvec{1}{0}^{\otimes n}
          + \beta' \tbcolvec{0}{1}^{\otimes n}\right).
  \]
  Let $T = \tbmatrix{a}{b}{b}{-a}$ for $a, b \in \mathbb{C}$ such that $a^2 + b^2 = 1$.
  Then
  \[
           \tbcolvec{a+b}{b-a}^{\otimes n}
   + \beta \tbcolvec{a-b}{a+b}^{\otimes n}
   = \lambda \left(\tbcolvec{1}{0}^{\otimes n}
          + \beta' \tbcolvec{0}{1}^{\otimes n}\right).
  \]
  By Lemma~\ref{lem:vector:uniq} we have either $a - b = 0$ or $a + b = 0$. 
  Together with $a^2 + b^2 = 1$,
  the only solutions are $T = \pm \frac{1}{\sqrt{2}} \tbmatrix{1}{1}{1}{-1}$ or
  $T = \pm \frac{1}{\sqrt{2}} \tbmatrix{1}{-1}{-1}{-1} = \pm \frac{1}{\sqrt{2}} \tbmatrix{1}{1}{1}{-1} \tbmatrix{0}{-1}{1}{0}$.
  Since $\pm \frac{1}{\sqrt{2}} I_2, \pm \frac{1}{\sqrt{2}} \tbmatrix{0}{-1}{1}{0} \in \StabP$, this case is complete.
  \item Suppose $\mathcal{F} \subseteq H' \tbmatrix{1}{1}{i}{-i} \mathscr{P}$.
  Then $g \in \tbmatrix{1}{1}{i}{-i} \mathscr{P}$, and $\theta(g) = \theta(\tbcolvec{1}{1}, \tbcolvec{1}{-1}) = 0$ by Lemma~\ref{lem:symmetric:angle:AP-trans}.

  However, any $h\in \tbmatrix{1}{1}{i}{-i} \mathscr{P}$ that is non-degenerate and has arity at least $3$
  must have the form $c\tbcolvec{1}{i}^{\otimes n} + d\tbcolvec{1}{-i}^{\otimes n}$ for some nonzero $c, d \in \mathbb{C}$, 
  which implies that $\theta(h) = -1$.
  This contradicts $\theta(g) = 0$.
  \qedhere
 \end{enumerate}
\end{proof}

\begin{lemma} \label{lem:trans:P2}
 Let $\mathcal{F}$ be a set of symmetric signatures and
 suppose $\mathcal{F}$ contains a non-degenerate signature $f \in \mathscr{P}_2$ of arity $n \ge 3$.
 Then $\mathcal{F}$ is $\mathscr{P}$-transformable
 iff all non-degenerate signatures in $\mathcal{F}$ are contained in $\mathscr{P}_2 \union \{=_2\}$.
\end{lemma}

\begin{proof}
 Suppose $\mathcal{F}$ is $\mathscr{P}$-transformable.
 Let $Z = \frac{1}{\sqrt{2}} \tbmatrix{1}{1}{i}{-i}$.
 Then by Lemma~\ref{lem:product:trans:so2},
 $\mathcal{F} \subseteq Z \mathscr{P}$ or
 there exists an $H \in \mathbf{SO}_2(\mathbb{C})$ such that $\mathcal{F} \subseteq H \mathscr{P}$.
 In first case, we show that all the non-degenerate symmetric signatures in $Z \mathscr{P}$ are contained in $\mathscr{P}_2 \union \{=_2\}$.
 Then we show that the second case is impossible.
 \begin{enumerate}
  \item Suppose $\mathcal{F} \subseteq Z \mathscr{P}$.
  Let $g \in Z \mathscr{P}$ be a symmetric non-degenerate signature of arity $m$.
  If $(Z^{-1})^{\otimes 2} g = \lambda [0,1,0]$ is the binary disequality signature up to a nonzero scalar $\lambda \in \mathbb{C}$,
  then
  \begin{align*}
   g
   = \lambda Z^{\otimes 2} \left(\begin{smallmatrix} 0 \\ 1 \\ 1  \\ 0 \end{smallmatrix}\right)
   = \lambda  \left(\begin{smallmatrix} 1 \\ 0 \\ 0  \\ 1 \end{smallmatrix}\right)
  \end{align*}
  is the binary equality signature $=_2$.
  Otherwise, we can express $g$ as
  \begin{align*}
   g
   &= c Z^{\otimes m} \left(\tbcolvec{1}{0}^{\otimes m}
                    + \beta \tbcolvec{0}{1}^{\otimes m}\right)\\
   &= c \left(\tbcolvec{1}{ i}^{\otimes m}
      + \beta \tbcolvec{1}{-i}^{\otimes m}\right)
  \end{align*}
  for some $c, \beta \neq 0$ with $m \ge 2$.
  Thus, $g \in \mathscr{P}_2 = \mathscr{A}_2$ by Lemma~\ref{lem:single:P2}.
  We conclude that the symmetric non-degenerate subset of $Z \mathscr{P}$ is contained in $\mathscr{P}_2 \union \{=_2\}$.
  Therefore, the non-degenerate subset of $\mathcal{F}$ is contained in $\mathscr{P}_2 \union \{=_2\}$.
  \item Suppose $\mathcal{F} \subseteq H \mathscr{P}$.
    By assumption, $\mathcal{F}$ contains $f \in \mathscr{P}_2 = \mathscr{A}_2$ of arity $n\ge 3$. 
  After a suitable normalization by a scalar, we have
  \[
   f = \tbcolvec{1}{i}^{\otimes n} + \beta \tbcolvec{1}{-i}^{\otimes n}
  \]
  for some $\beta \neq 0$ by Lemma~\ref{lem:single:P2}.
  Let $g = (H^{-1})^{\otimes n}f$ so that
  \[
   g = \left(H^{-1}\right)^{\otimes n} \left(\tbcolvec{1}{ i}^{\otimes n}
                                     + \beta \tbcolvec{1}{-i}^{\otimes n}\right).
  \]
  In particular, $f$ and $g$ have the same arity $n \ge 3$.
  By Lemma~\ref{lem:symmetric:angle:AP-trans},
  $\theta(g) = \theta(\tbcolvec{1}{i}, \tbcolvec{1}{-i}) = -1$ since $H^{-1} \in \mathbf{O}_2(\mathbb{C})$.
  However, $g \in \mathscr{P}$ must be of the form $\tbcolvec{c}{0}^{\otimes n} + \tbcolvec{0}{d}^{\otimes n}$ for some nonzero $c, d \in \mathbb{C}$, 
  which has $\theta(g) = 0$.
  This is a contradiction.
 \end{enumerate}

 It is easy to see that all of above is reversible.
 Therefore sufficiency follows.
\end{proof}

Now we are ready to show how to decide if a finite set of signatures is $\mathscr{P}$-transformable.
To avoid trivialities, we assume $\mathcal{F}$ contains a non-degenerate signature of arity at least~$3$.
If every non-degenerate signature in $\mathcal{F}$ has arity at most two, then $\Holant(\mathcal{F})$ is tractable.

\begin{theorem}
 There is a polynomial-time algorithm to decide,
 for any finite input set $\mathcal{F}$ of symmetric signatures containing a non-degenerate signature $f$ of arity $n \geq 3$,
 whether $\mathcal{F}$ is $\mathscr{P}$-transformable.
\end{theorem}

\begin{proof}
 By Lemma~\ref{lem:single:product},
 we can decide if $f$ is in $\mathscr{P}_k$ for some $k \in \{1,2\}$.
 If not, then by Lemma~\ref{lem:cha:product}, $\mathcal{F}$ is not $\mathscr{P}$-transformable.
 Otherwise, $f \in \mathscr{P}_k$ for some unique $k$.
 Depending on $k$,
 we apply Lemma~\ref{lem:trans:P1} or Lemma~\ref{lem:trans:P2} to check if $\mathcal{F}$ is $\mathscr{P}$-transformable.
\end{proof}